%% file: ChPTtheta.tex
\documentclass[aps,prd,preprint,nofootinbib,superscriptaddress]{revtex4}

\usepackage{graphicx}
\usepackage{psfrag}

\begin{document}
\preprint{UTHEP-585}
\title{Chiral perturbation theory in a $\theta$ vacuum}

\newcommand{\Tsukuba}{
  Graduate School of Pure and Applied Sciences, University of Tsukuba,
  Tsukuba 305-8571, Japan
}
\newcommand{\BNL}{
  Riken BNL Research Center, Brookhaven National Laboratory, Upton,
  New York 11973, USA
}

\newcommand{\Nagoya}{
  Department of Physics, Nagoya University, Nagoya 464-8602, Japan
}

\author{Sinya Aoki}
\affiliation{\Tsukuba}
\affiliation{\BNL}
\author{Hidenori Fukaya}
\affiliation{\Nagoya}

\begin{abstract}
We consider chiral perturbation theory with a
nonzero $\theta$ term.
Because of the {\it CP} violating term, the vacuum of chiral
fields is shifted to a nontrivial element 
on the $SU(N_f)$ group manifold.
%In the chiral Lagrangian, this shift can be 
%regarded as a source which 
The {\it CP} violation also provides mixing
of different {\it CP} eigenstates, 
between scalar and pseudoscalar,
or vector and axialvector operators.
We investigate upto ${\cal O}(\theta^2)$ effects 
on the mesonic two-point correlators of 
chiral perturbation theory to the one-loop order.
We also address the effects of fixing topology, 
by using saddle-point integration in the
Fourier transform with respect to $\theta$.
\end{abstract}

\maketitle

\input{sec1.tex}

\input{sec2.tex}
\input{sec3.tex}

\input{sec4.tex} 
\input{sec5.tex}
\input{sec5-2.tex}
\input{sec6.tex}
\input{appendix.tex}
\input{appendix2.tex}

\input{ref.tex}
\end{document}

%% file: sec1.tex
%%%%%%%%%%%%%%%%%%%%%%%%%%%%%%%%%%%%%%%%%%%%%%%%%%%
%%%%%%%%%%%%%%%%%%%%%%%%%%%%%%%%%%%%%%%%%%%%%%%%%%%
%%%%%%%%%%%%%%%%%%%%%%%%%%%%%%%%%%%%%%%%%%%%%%%%%%%
\section{Introduction}
\label{sec:intro}
%\setcounter{equation}{0}
%%%%%%%%%%%%%%%%%%%%%%%%%%%%%%%%%%%%%%%%%%%%%%%%%%%a
%%%%%%%%%%%%%%%%%%%%%%%%%%%%%%%%%%%%%%%%%%%%%%%%%%%
%%%%%%%%%%%%%%%%%%%%%%%%%%%%%%%%%%%%%%%%%%%%%%%%%%%
 
The low energy limit of quantum chromo dynamics 
(QCD) is full of nonperturbative phenomena 
such as quark confinement and chiral symmetry breaking.
%Due to its non-linearity and strong coupling of 
%interactions 
It has, however, been very difficult to 
analytically investigate these phenomena 
from the first principle due to 
its nonlinearity and strong coupling of interactions. 
Lattice QCD  \cite{Aubin:2004fs}
and chiral perturbation
theory (ChPT) \cite{Weinberg:1968de, Gasser:1983yg}
have played prominent roles in studying 
such dynamical phenomena of QCD.
The nonperturbative calculations of lattice QCD
can be numerically performed utilizing the latest
computational resources, while 
ChPT, an effective theory of pions,
allows us to perturbatively 
treat the very low energy limit of QCD.
They are complementary each other and 
have mutually developed. 
Lattice QCD in principle can determine the
low energy constants of ChPT, some of which are difficult to determine from experimental inputs. On the other hand,
ChPT provides a theoretical guideline of how to
extrapolate the lattice data to the 
near-zero quark mass limit (chiral extrapolation) 
or the infinite volume limit (finite size scaling).

An interesting extension of QCD is to introduce
a {\it CP} violation term, known as the $\theta$ term.
Since it is written as a total derivative in the
QCD Lagrangian, it exists only when the gauge fields 
can have a nontrivial topological structure,
or winding numbers.
By partial integration, this $\theta$ term can be
regarded as the phase of superposition of 
different vacua in the Hamilton picture, 
which we call the $\theta$ vacuum \cite{Vicari:2009}.

{\it CP} is invariant non only at $\theta=0$ but also 
at $\theta=\pi$. 
It is, however, believed that {\it CP} is spontaneously
broken at $\theta=\pi$, where the theory has 
two {\it CP} violating vacua \cite{Dashen:1970et}.
Moreover, it is expected that there exists a first-order
phase transition in $0<\theta\leq \pi$.
These issues have been investigated mainly
using effective theories \cite{Witten:1980sp}.

It is little known in nature why this {\it CP} violation term 
is invisibly suppressed, which is the so-called strong {\it CP} problem.
The neutron electric dipole moment
has not been observed in the experiments, from which
one can estimate $|\theta | \lesssim 10^{-10}$ \cite{NEDM}.
%Although the question is beyond QCD itself, 
The lattice QCD community has also tried 
to {\it quantify}  the strong {\it CP} problem \cite{Aoki:1990}.
It remains, however, to be one of the most difficult problems 
in lattice QCD, since the $\theta$ term gives a complex action, which
leads to the notorious phase problem in Monte-Carlo simulations.
%in lattice QCD: the
%phase problem of Monte Carlo studies since
%the $\theta$ term makes the path-integration complex.
With various elaborated approaches, many groups have 
investigated the $\theta$ vacuum in the lattice 
simulations \cite{Azcoiti:2002vk, Fukaya:2004kp}.

There exists another motivation in studying the $\theta$ vacuum of QCD.
One can consider a fixed topological sector of the theory,
by Fourier transform of the partition function 
with respect to $\theta$.
%Using the saddle point expansion, 
Expanding the vacuum energy as $f(\theta)=c_2\theta^2/2+c_4\theta^4/4!\cdots$
and performing the saddle-point integration order by order,
one can investigate
how the topology affects the physical quantities
and evaluate the difference between the physics 
between the $\theta$ vacuum and the one 
at fixed topology \cite{Brower:2003yx, Aoki:2007ka}.
Inversely, it is also possible to determine the vacuum energy
of QCD as a power series of $\theta$, 
from the physical observables at fixed topology.
In Refs.\cite{Brower:2003yx, Aoki:2007ka}, a general formula
to ${\cal O}(1/V^2)$ is calculated which converts 
the observables at fixed topology to those in $\theta=0$ vacuum
treating $c_2$, $c_4,\cdots$ as unknown parameters.
It is, therefore, an important task to calculate
these parameters within QCD or the low energy effective theory
to quantify the effects of fixing topology.

In fact, QCD at fixed topology can be investigated
by lattice QCD simulations.  
Employing the overlap Dirac operator \cite{Neuberger:1997fp}
which preserves the exact chiral symmetry,
we are able to define the topological charge on the lattice.
JLQCD and TWQCD Collaborations are using a topology 
preserving way \cite{Izubuchi:2002pq}
to avoid discontinuities
of the overlap fermion determinant,
which considerably reduces the computational cost.
%Since the exact chiral symmetry of the Dirac operator 
%provides a clean way in comparing the lattice data
%to ChPT, they have been successful 
The conversion formula between $\theta$ vacuum and fixed topology
is, therefore, essential in extracting
the low energy constants such as the chiral condensate $\Sigma$ \cite{Fukaya:2007fb},
the pion decay constant $F$ \cite{Fukaya:2007pn}
and the topological susceptibility \cite{Aoki:2007pw}.\\

Because of its {\it global}  topological nature,
the $\theta$ vacuum effect is totally infra red physics
and, therefore, should be described by the
lightest particles, or pions, 
within chiral perturbation theory.
In this paper, we discuss ChPT with nonzero value of $\theta$
as well as at a fixed topology.
The formulation and general qualitative discussion
are already given in Refs. \cite{Gasser:1983yg,e-regime}.
Our goal is to explicitly calculate the meson correlators to 
the next-to-leading order (NLO)
in the $\theta$ vacuum (and in a fixed topological sector)
at finite volume, which may be directly compared with lattice QCD. 
In Sec.~\ref{sec:LO} we will observe that the vacuum of chiral
fields is located not at identity but at 
a nontrivial element on the $SU(N_f)$ group manifold. 
In the $p$ expansion of the chiral Lagrangian, 
this vacuum shift provides as a source 
which mixes the different {\it CP} eigenstates 
(Sec.~\ref{sec:NLO}).
In Sec.~\ref{sec:2pt}, 
we will calculate the $\theta$ vacuum effect
upto ${\cal O}(\theta^2)$
on the mesonic two-point correlation functions 
in ChPT to NLO.
In Sec.~\ref{sec:fixedQ} we will also address the physics at fixing topology, 
by using the saddle-point integration in the Fourier transform with respect to $\theta$,
as discussed above.
The concluding remarks are given in Sec.~\ref{sec:conclusion}.

%% file: sec2.tex
%%%%%%%%%%%%%%%%%%%%%%%%%%%%%%%%%%%%%%%%%%%%%%%%%%%
%%%%%%%%%%%%%%%%%%%%%%%%%%%%%%%%%%%%%%%%%%%%%%%%%%%
%%%%%%%%%%%%%%%%%%%%%%%%%%%%%%%%%%%%%%%%%%%%%%%%%%%
\section{Chiral Lagrangian to the leading order}
\label{sec:LO}
%\setcounter{equation}{0}
%%%%%%%%%%%%%%%%%%%%%%%%%%%%%%%%%%%%%%%%%%%%%%%%%%%a
%%%%%%%%%%%%%%%%%%%%%%%%%%%%%%%%%%%%%%%%%%%%%%%%%%%
%%%%%%%%%%%%%%%%%%%%%%%%%%%%%%%%%%%%%%%%%%%%%%%%%%%

We consider the $N_f$-flavor chiral Lagrangian
in the $\theta$ vacuum,
%%%%%%%%%%%%%%% Eq ChPT Lagrangian %%%%%%%%%%%%%%%
\begin{eqnarray}
\mathcal{L}
&=&
\frac{F^2}{4}{\rm Tr}
[\partial_\mu U(x)^\dagger\partial_\mu U(x) ]
-\frac{\Sigma}{2}{\rm Tr}
[\mathcal{M}^{\dagger}e^{-i\theta /N_f}U(x)
+U(x)^\dagger e^{i\theta /N_f}\mathcal{M}]+\cdots,
\end{eqnarray}
where $U(x)\in SU(N_f)$, $\Sigma$ is the chiral condensate,
and $F$ denotes the pion decay constant 
both in the chiral limit.
Here $\theta$, the QCD vacuum angle, 
appears as the phase of the mass term,
reflecting the picture that the $\theta$ term
can be converted to the chiral rotation of the 
quark bilinears through the anomalous Ward-Takahashi identity. 
In the mass matrix,
%%%%%%%%%%%%%%% Eq mass matrix %%%%%%%%%%%%%%%%%%%
\begin{eqnarray}
\mathcal{M} &=& 
\mathrm{diag}(\underbrace{m_{v},m_{v^\prime},\cdots}_{N_v},
\underbrace{m_1,m_2,\cdots}_{N_f}),
\end{eqnarray}
we have $N_v$ valence flavors and $N_f$ dynamical flavors.

In the partially quenched case \cite{Sharpe:1997by},
we use the so-called replica trick,
where the calculations are done with 
[$N_f+N_v+(N-N_v)$]-flavor theory and then 
the replica limit $N \to 0$ is taken.
The full theory results are precisely obtained by choosing
$m_v=m_f$ where $m_f$ denotes one of the dynamical quark masses.

The system is assumed to be in the so-called $p$ regime,
where the Euclidean space-time volume $V=L^3 T$ is 
large enough so that the perturbative expansion is 
performed according to the counting rule,
\begin{eqnarray}
\partial_\mu \sim {\cal O}(p),\;\;\;
\xi(x) \sim {\cal O}(p),\;\;\;
\mathcal{M}\sim {\cal O}(p^2),\;\;\;
T, L \sim {\cal O}(1/p),
\end{eqnarray}
in the units of a cutoff scale.\\

Note that due to the vacuum angle $\theta$,
the expectation value of $U(x)$ is located not
at the identity but a nontrivial element of $SU(N_f)$ 
(let us denote $U_0$). 
It is, therefore, useful to define the new variable
and mass matrix,
%%%%%%%%%%%%% Eq reparametrization %%%%%%%%%%%%%%%%%
\begin{eqnarray}
U(x)\equiv U_0\tilde{U}(x), &
\mathcal{M}_\theta \equiv U^\dagger_0 e^{i\theta /N_f}\mathcal{M},
\end{eqnarray}
and rewrite the Lagrangian,
%%%%%%%%%%%%% Eq Lagrangian rewritten %%%%%%%%%%%%%%
\begin{eqnarray}
\label{eq:ChPTLO}
\mathcal{L}
&=&
\frac{F^2}{4}{\rm Tr}
[\partial_\mu \tilde{U}(x)^\dagger\partial_\mu \tilde{U}(x) ]
-\frac{\Sigma}{2}{\rm Tr}
[\mathcal{M}^{\dagger}_\theta \tilde{U}(x)
+\tilde{U}(x)^\dagger\mathcal{M}_\theta]+\cdots,
\end{eqnarray}
where $\tilde{U}(x)$ can be expanded around the identity as usual;
%%%%%%%%%%%%% Eq expansion %%%%%%%%%%%%%%%%%%%%%%%%%
\begin{eqnarray}
\label{eq:Uexp}
\tilde{U}(x)=\exp\left(i\frac{\sqrt{2}\xi(x)}{F}\right)
\sim 1 + i\frac{\sqrt{2}\xi(x)}{F} 
-\frac{\xi^2(x)}{F^2}-\frac{i\sqrt{2}\xi^3(x)}{3F^3}
+\frac{\xi^4(x)}{6F^4}
+\cdots,
\end{eqnarray}
where $\xi$ is an element of $SU(N_f)$ Lie algebra.
%Since we only treat $\tilde{U}(x)$ in this work,
%the tilde is omitted in the following for simplicity.\\

By this vacuum shift, Eq.(\ref{eq:ChPTLO}) explicitly shows
that the nonzero $\theta$ vacuum physics is equivalent
to that in the $\theta=0$ vacuum but with a 
complex mass matrix $\mathcal{M}_\theta$.
Our task in the following is, therefore, 
to determine $U_0$ (or equivalently $\mathcal{M}_\theta$)
and then to calculate the difference between 
the systems with the complex $\mathcal{M}_\theta$ and
a simply real diagonal $\mathcal{M}$.

%%%%%%%%%%%%%%%%%%%%%%%%%%%%%%%%%%%%%%%%%%%%%%%%
\subsection{Vacuum shift $U_0$}
%%%%%%%%%%%%%%%%%%%%%%%%%%%%%%%%%%%%%%%%%%%%%%%%

Let us first calculate the vacuum expectation value $U_0$,
which minimizes the Lagrangian density of the zero-mode
%%%%%%%%%%%% Eq zero-mode Lagrangian %%%%%%%%%%%
\begin{eqnarray}
\label{eq:L0}
\mathcal{L}_0=-\frac{\Sigma}{2}{\rm Tr}
[\mathcal{M}^{\dagger}e^{-i\theta /N_f}U_0
+U^\dagger_0 e^{i\theta /N_f}\mathcal{M}].
\end{eqnarray}
For small $\theta$, %\sim {\cal O}(p)$ (in the lattice units), 
by parametrizing $U_0=\exp(i\xi^0)$, 
the problem is equivalent to finding 
the minimum of a potential
%%%%%%%%%%%% Eq potential for xi^0 %%%%%%%%%%%%%
\begin{eqnarray}
V(\xi^0)=\sum_i^{N_f} m_i\left[- \sin \left(\frac{\theta}{N_f}\right)
\left[\xi^0-\frac{1}{6}(\xi^0)^3\right]_{ii}
+\frac{1}{2}\cos \left(\frac{\theta}{N_f}\right)
\left[(\xi^0)^2-\frac{1}{12}(\xi^0)^4\right]_{ii}\right]
+\lambda \mathrm{Tr}[\xi^0],
\end{eqnarray}
where $\lambda$ denotes the Lagrange's multiplier
to guarantee the traceless solution.

It can be recursively shown, order by order,
that all the off-diagonal parts of $\xi^0$ are zero.
% since
%%%%%%%%%%%% Eq off-diagonal xi^0 %%%%%%%%%%%%%%
%\begin{equation}
%\frac{\partial V(\xi^0)}{\partial \xi^0_{ij}}
%=\frac{1}{2}\cos \left(\frac{\theta}{N_f}\right)
%(m_i+m_j)[\xi^0]_{ji}=0\;\;\;\mbox{for}(i\neq j),
%\end{equation}
%at the minimum.
For the diagonal elements, 
%solving the equations
%%%%%%%%%%%% Eq diagonal xi^0 %%%%%%%%%%%%%%%%%%
%\begin{eqnarray}
%\frac{\partial V(\xi^0)}{\partial \xi^0_{ii}}
%&=&-\sin \left(\frac{\theta}{N_f}\right)m_i
%\left[1-\frac{1}{2}(\xi^0_{ii})^2\right]
%+\cos \left(\frac{\theta}{N_f}\right)
%m_i\left[\xi^0_{ii}-\frac{1}{6}(\xi^0_{ii})^3\right]+\lambda=0,\\
%\frac{\partial V(\xi^0)}{\partial \lambda}
%&=&0,
%\end{eqnarray}
one obtains
%%%%%%%%%%%% Eq xi0 solution %%%%%%%%%%%%%%%%%%%
\begin{eqnarray}
\label{eq:xi0}
\xi^0_{ii}&=&\bar{\theta}-\frac{\bar{m}}{m_i}\theta
+a_i\theta^3 +{\cal O}(\theta^5),
\end{eqnarray}
where 
\begin{eqnarray}
\bar{\theta}\equiv \frac{\theta}{N_f}, \;\;\;\;\;
\bar{m}\equiv \frac{1}{\sum_f^{N_f}1/m_f},\;\;\;\;\;
%\end{eqnarray}  
%and
%%%%%%%%%%%% Eq a_i
%\begin{eqnarray}
a_i \equiv \frac{\bar{m}^3}{6}
\left(\frac{\bar{m}}{m_i}\sum_f^{N_f}\frac{1}{m_f^3}-\frac{1}{m_i^3}\right).
\end{eqnarray}
Note that the partially quenched result is obtained 
simply by taking $m_i=m_v$ after the replica limit.
%In the full theory limit $m_v\to m_f$,
%$\xi^0_{vv}$ agrees with $\xi^0_{ff}$.
It is also notable that $\xi^0_{ii}=0$ when $m_i$'s are
all degenerate.

%Defining 
%%%%%%%%%%%% Eq theta_i
%\begin{eqnarray}
%\theta_i &\equiv& \frac{\bar{m}}{m_i}\theta-a_i\theta^3,
%\end{eqnarray}
Now the original Lagrangian in Eq.(\ref{eq:L0}) 
is greatly simplified;
%%%%%%%%%%%% Eq L_0 simplified %%%%%%%%%%%
\begin{eqnarray}
\mathcal{L}_0 &=& -\Sigma 
\left[\sum_f^{N_f}m_f \cos \left(\frac{\bar{m}}{m_f}\theta-a_f\theta^3\right) \right],
\end{eqnarray}
from which one can read off the vacuum energy density
at the tree level
\footnote{
For the $N_f=2$ case, a nonperturbative expression of 
the vacuum energy density
is known \cite{Lenaghan:2001ur},
from which one can read off $\chi_t$, $c_4$  \cite{Mao:2009sy},
and any higher order coefficients $c_{2n}$'s.
}, as a function of $\theta$.
In particular, the coefficients of $\theta^2$ (the
topological susceptibility $\chi_t$)  
and $\theta^4$ (we denote $c_4$),
%%%%%%%%%%%% Eq chi_t, c_4 %%%%%%%%%%%%%%%
\begin{eqnarray}
\chi_t &=& \left. \frac{d^2 \mathcal{L}_0}{d\, \theta^2}\right\vert_{\theta=0}=\bar{m}\Sigma,\\
\label{eq:c4tree}
c_4&=& \left. \frac{d^4 \mathcal{L}_0}{d\, \theta^4}\right\vert_{\theta=0}=  -\bar{m}\Sigma \left(\sum_f^{N_f}\frac{\bar{m}^3}{m_f^3}\right),
\end{eqnarray}
are important when we consider the effect of fixing topology
as previously discussed in Refs.~\cite{Brower:2003yx, Aoki:2007ka}.

Another important observation follows from the fact
%%%%%%%%%%% Eq Im M %%%%%%%%%%%%%%%%%%%%%%%%%%%%
\begin{eqnarray}
\mathrm{Im}(\mathcal{M}_\theta)
%&=& \frac{\theta}{\sum^{N_f}_f1/m_f} {\bf 1} +{\cal O}(p^4)
%\nonumber\\
%&=&\frac{m_l m_h \theta}{N_l m_h + N_h m_l} {\bf 1} +{\cal O}(p^4)
%\nonumber\\
&=& \bar{m} \theta {\bf 1} +{\cal O}(\theta^3), 
\end{eqnarray}
where ${\bf 1}$ denotes the $N_f\times N_f$ identity matrix.
%\begin{eqnarray}
%\label{eq:mprime}
%\bar{m} (\theta)&\equiv&  \frac{\theta}{\sum^{N_f}_f1/m_f}. %{\bf 1}
%%= \frac{m_l m_h \theta}{N_l m_h + N_h m_l}, %{\bf 1},
%\end{eqnarray}
Noting ${\rm Tr}\xi =0$, the contribution from the imaginary part of 
the mass matrix then becomes $\sim {\cal O}(p^5)$
%%%%%%%%%%% Eq U-U^dag %%%%%%%%%%%%%%%%%%%%%%%%%%
%\begin{eqnarray}
%\frac{i\Sigma}{2}{\rm Tr}[\mathrm{Im}(\mathcal{M}_\theta) 
%(\tilde{U}(x)-\tilde{U}(x)^\dagger)]=
%\frac{\sqrt{2}\bar{m}\theta\Sigma}{3F^3}{\rm Tr}[\xi^3(x)]\sim {\cal O}(p^5),
%\end{eqnarray}
so that we can neglect it at the leading order.
\footnote{It of course gives contributions at NLO.}
%since the odd number of $\xi$'s
%need to be pulled down from the action twice, at least.
%Therefore, the imaginary part of the mass term in the chiral Lagrangian 
%can be ignored at the leading order.
%%%%%%%%%%% Eq mass term after U_0 determined %%%%%%%%%%%%
%\begin{eqnarray}
%-\frac{\Sigma}{2}{\rm Tr}
%[\mathrm{Re}(\mathcal{M}_\theta) 
%(\tilde{U}(x)+\tilde{U}(x)^\dagger)].
%\end{eqnarray}

In this subsection, we have derived the vacuum expectation
value $U_0$ upto ${\cal O}(\theta^4)$ level.
The most part of this paper, however, requires only 
${\cal O}(\theta^2)$ contribution and one can 
neglect the 3rd term of Eq.(\ref{eq:xi0}) or set $a_i=0$.

%%%%%%%%%%%%%%%%%%%%%%%%%%%%%%%%%%%%%%%%%%%%%%%%
\subsection{Propagator of $\xi(x)$}
%%%%%%%%%%%%%%%%%%%%%%%%%%%%%%%%%%%%%%%%%%%%%%%%

Let us now expand the Lagrangian in $\xi$,
%%%%%%%%% Eq LO Lagrangian expanded %%%%%%%%%%%%%
\begin{eqnarray}
\label{eq:Lagrangian}
\mathcal{L}&=&
%\frac{F^2}{4}{\rm Tr}
%[\partial_\mu \tilde{U}(x)^\dagger\partial_\mu \tilde{U}(x) ]
%-\frac{\Sigma}{2}{\rm Tr}
%[\mathrm{Re}(\mathcal{M}_\theta) 
%(\tilde{U}(x)+\tilde{U}(x)^\dagger)]+\cdots,\nonumber\\
%&=& \frac{1}{2}{\rm Tr}(\partial_\mu \xi(x))^2
%+\frac{\Sigma}{F^2}\sum_i^{N_f}({\rm Re}\mathcal{M}_\theta)_{ii}
%[\xi^2(x)]_{ii} +\cdots%{\cal O}(p^6),
%\nonumber\\
%&=&
\mathcal{L}_0+\frac{1}{2}\sum_{i,j}^{N_f}[\xi(x)]_{ij}
(-\partial_\mu^2 + M^2_{ij}(\theta))[\xi(x)]_{ji} +\cdots%{\cal O}(p^6).
\end{eqnarray}
Here $M^2_{ij}(\theta)$ is given by
%%%%%%%%% Eq M_ij definition %%%%%%%%%%%%%%%%%%%%
\begin{eqnarray}
M^2_{ij}(\theta) &=& \frac{\Sigma}{F^2}
\left(m_i(\theta)+m_j(\theta)\right),
\end{eqnarray}
where $m_i(\theta)$ is defined by
%%%%%%%%% Eq m_i(\theta) %%%%%%%%%%%%%%%%%%%%%%%%%
\begin{eqnarray}
m_i(\theta)&=& m_i\cos\left(\frac{\bar{m}}{m_i}\theta-a_i\theta^3\right) =
m_i \left(
%\cos\left(
1-\frac{1}{2}\frac{\bar{m}^2}{m_i^2}\theta^2 \right)+{\cal O}(\theta^4),
%\frac{\theta}{m_i(\sum_f^{N_f}\frac{1}{m_f})}\right),
%\end{eqnarray}
%where
%\begin{eqnarray}
%\;\;\; \theta_i \equiv \frac{\theta}{m_i(\sum_f^{N_f}\frac{1}{m_f})},
\end{eqnarray}
and again there is no significant difference 
even if we extend the theory to the partially quenched one;
we just set $m_i=m_v$.
%The valence mass is given by,
%%%%%%%%% Eq m_v(\theta) %%%%%%%%%%%%%%%%%%%%%%%%%
%\begin{eqnarray}
%m_v(\theta)&=&m_v \left(
%\cos\left(
%1-\frac{\theta_v^2}{2}\right),
%\frac{\theta}{m_i(\sum_f^{N_f}\frac{1}{m_f})}\right),
%\end{eqnarray}
%where
%\begin{eqnarray}
%\;\;\; \theta_v \equiv \frac{\theta}{m_v(\sum_f^{N_f}\frac{1}{m_f})},
%\end{eqnarray}

%or more explicitly denoting the flavor index,
%%%%%%%%% Eq m_l(\theta), m_v, m_h %%%%%%%%%%%%%%%%%%%%%%%%%
%\begin{eqnarray}
%m_v(\theta)&=&m_v (1-\theta_v^2/2),
%\;\;\;\theta_v = \frac{m_l m_h\theta}{m_v(N_lm_h+N_hm_l)},
%%\right),\\
%\\
%m_l(\theta)&=&m_l(1-\theta_l^2/2),
%\;\;\; %\cos\left(
%\theta_l = \frac{m_h\theta}{N_lm_h+N_hm_l},\\
%%\right),\\
%m_h(\theta)&=&m_h(1-\theta_h^2/2),
%\;\;\;%\cos\left(
%\theta_h = \frac{m_l\theta}{N_lm_h+N_hm_l}.
%\end{eqnarray}

The Feynmann propagator (in a finite volume) is then obtained:
%%%%%%%%% Eq m_l(\theta), m_v, m_h %%%%%%%%%%%
\begin{eqnarray}
\langle \xi_{ij}(x)\xi_{kl}(y)\rangle_\xi
&=&\delta_{il}\delta_{jk}\Delta(x-y, M^2_{ij}(\theta))
-\delta_{ij}\delta_{kl}G(x-y,M^2_{ii}(\theta),M^2_{kk}(\theta)),
%\nonumber\\
\end{eqnarray}
where the second term comes from the traceless
constraint ${\rm Tr}\xi=0$.
The definitions of $\Delta$ and $G$ are given by
(unless $N_f = 0$)
%%%%%%%%% Eq Delta ,G %%%%%%%%%%%%%%%%%%%%%%%%%
\begin{eqnarray}
\Delta(x,M^2)&=&\frac{1}{V}\sum_{p}\frac{e^{ipx}}{p^2+M^2},\\
G(x,M_{ii}^2,M^2_{jj})&=&
\frac{1}{V}\sum_{p}\frac{e^{ipx}}{(p^2+M_{ii}^2)(p^2+M_{jj}^2)
\left(\sum^{N_f}_f\frac{1}{p^2+M^2_{ff}(\theta)}
%+\frac{N_h}{p^2+M^2_{hh}(\theta)}
\right)},
\label{eq:propG}
%\nonumber\\
\end{eqnarray}
where the summation is taken over the 4-momentum
%%%%%%%%% Eq Sum_p %%%%%%%%%%%%%%%%%%%%%%%%%%%%
\begin{eqnarray}
p=2\pi(n_t/T, n_x/L, n_y/L, n_z/L),
\end{eqnarray}
with integers $n_i$'s.

%% file: sec3.tex
%%%%%%%%%%%%%%%%%%%%%%%%%%%%%%%%%%%%%%%%%%%%%%%%%%%
%%%%%%%%%%%%%%%%%%%%%%%%%%%%%%%%%%%%%%%%%%%%%%%%%%%
%%%%%%%%%%%%%%%%%%%%%%%%%%%%%%%%%%%%%%%%%%%%%%%%%%%
\section{One-loop corrections}
\label{sec:NLO}
%\setcounter{equation}{0}
%%%%%%%%%%%%%%%%%%%%%%%%%%%%%%%%%%%%%%%%%%%%%%%%%%%a
%%%%%%%%%%%%%%%%%%%%%%%%%%%%%%%%%%%%%%%%%%%%%%%%%%%
%%%%%%%%%%%%%%%%%%%%%%%%%%%%%%%%%%%%%%%%%%%%%%%%%%%

Since we are interested in at most one-loop corrections
to the two-point functions,
we can take, in advance, some Wick contractions
in ${\cal O}(p^5)$ or ${\cal O}(p^6)$ contributions
in the expansion of the leading order 
Lagrangian Eq.(\ref{eq:Lagrangian}).
% which results 
%in the correction to the kinetic term or 
%mass term in the Lagrangian.

Before performing the one-loop calculation, we here introduce the NLO terms 
of the chiral Lagrangian. 
Without source terms, 
we have 8 additional NLO terms 
whose low-energy constants are denoted by $L_i$'s ($i=1 \cdots 8$)\footnote{
When $N_f=2$, due to the pseudoreality, the number of independent terms 
is reduced to 5, of which coefficients are denoted by $l_i$'s 
\cite{Gasser:1983yg}.
}.
At ${\cal O}(p^5)$ and ${\cal O}(p^6)$, the terms with 
$L_1, L_2, L_3$ (and Wess-Zunimo-Witten term \cite{Wess:1971yu})
do not contribute to our calculation. 
\if0
and here we consider
\begin{eqnarray}
\label{eq:LNLO}
\mathcal{L}_{NLO}
%&=&
%\frac{F^2}{4}
&=&L_4\frac{2\Sigma}{F^2}{\rm Tr}[\partial_\mu \tilde{U}(x)^\dagger\partial_\mu \tilde{U}(x) ]
\times {\rm Tr}
[\mathcal{M}^{\dagger}_\theta \tilde{U}(x)
+\tilde{U}(x)^\dagger\mathcal{M}_\theta]
\nonumber\\
&&+L_5\frac{2\Sigma}{F^2}
{\rm Tr}[\partial_\mu \tilde{U}(x)^\dagger\partial_\mu \tilde{U}(x) 
%\times \frac{2\Sigma}{F^2}{\rm Tr}
(\mathcal{M}^{\dagger}_\theta \tilde{U}(x)
+\tilde{U}(x)^\dagger\mathcal{M}_\theta)]
\nonumber\\
&&-L_6 \left(\frac{2\Sigma}{F^2}
%{\rm Tr}[\partial_\mu \tilde{U}(x)^\dagger\partial_\mu \tilde{U}(x) 
%\times \frac{2\Sigma}{F^2}
{\rm Tr}
[\mathcal{M}^{\dagger}_\theta \tilde{U}(x)
+\tilde{U}(x)^\dagger\mathcal{M}_\theta]\right)^2
\nonumber\\
&&-L_7 \left(\frac{2\Sigma}{F^2}
%{\rm Tr}[\partial_\mu \tilde{U}(x)^\dagger\partial_\mu \tilde{U}(x) 
%\times \frac{2\Sigma}{F^2}
{\rm Tr}
[\mathcal{M}^{\dagger}_\theta \tilde{U}(x)
-\tilde{U}(x)^\dagger\mathcal{M}_\theta]\right)^2
\nonumber\\
&&-L_8 \left(\frac{2\Sigma}{F^2}\right)^2
{\rm Tr}
[\mathcal{M}^{\dagger}_\theta \tilde{U}(x)
\mathcal{M}^{\dagger}_\theta \tilde{U}(x)
+\tilde{U}(x)^\dagger\mathcal{M}_\theta
\tilde{U}(x)^\dagger\mathcal{M}_\theta].
\end{eqnarray}
\fi
For the terms with $L_4, L_5, L_6$, the NLO correction
is obtained in the same way as the $\theta=0$ case 
except for the change in the mass matrix;
$\mathcal{M} \to \mathrm{Re}\mathcal{M}_\theta$,
%which leads to terms in the effective Lagrangian,
%The correction to the valence sector kinetic term and
%the mass term from $L_4, L_5, L_6$ is immediately obtained,
while the $L_7$ and $L_8$ terms require a special care
of the imaginary part of the mass matrix,
$\mathrm{Im}\mathcal{M_\theta}=\bar{m}\theta {\bf 1}$. 

Expanding %Eq.(\ref{eq:LNLO}) we obtain
the chiral field according to Eq.(\ref{eq:Uexp}), we obtain
%%%%%%% Eq L_i corrections %%%%%%%%%%%%%%%%%%%%%
\begin{eqnarray}
\label{eq:LNLOex}
\mathcal{L}_{NLO}&=&
\sum_{i, j}\left[\frac{1}{2}%\sum^{N_v}_{v, j}
\partial_\mu \xi_{ij}
\partial_\mu \xi_{j i}
\right]\times \frac{8}{F^2}\left[
L_4\sum_f M^2_{ff}(\theta)+L_5M^2_{ij}(\theta)
\right]
\nonumber\\&&
+\sum_{i, j}\left[\frac{1}{2}\xi_{ij}\xi_{j i}\right]
\times M^2_{i j}(\theta)\frac{16}{F^2}
L_6\sum_fM^2_{ff}(\theta)%+L_8M^2_{ij}
\nonumber\\&&
+\frac{8L_7}{F^2}\sum^{N_f}_{i,j}M^2_{ii}(\theta)
M^2_{jj}(\theta)\xi_{ii}\xi_{jj}
+\frac{1}{2}\sum_{i,j}
\xi_{ij}\xi_{ji} %\times 
\frac{16L_8}{F^2}M^4_{ij}(\theta)
\nonumber\\&&
-\frac{8\sqrt{2}(N_fL_7+L_8)\bar{M}^2\theta}{F}
\sum_i^{N_f}M_{ii}^2(\theta)\xi_{ii}
%\nonumber\\&&
-\frac{16(N_fL_7+L_8)}{F^2}(\bar{M}^4\theta^2)
\sum^{N_f}_{i,j}\frac{1}{2}\xi_{ij}\xi_{ji},
\nonumber\\&&
-4L_6 (\sum_fM^2_{ff}(\theta))^2-2L_8\sum_fM^4_{ff}(\theta=0)
+4N_f(N_fL_7+L_8)\bar{M}^4\theta^2 +{\cal O}(p^7),
\end{eqnarray}
where $\bar{M}^2= 2 \bar{m}\Sigma/F^2$ and we have
used $[({\rm Re}\mathcal{M}_\theta)^2]_{ii}=
[(\mathcal{M}_{\theta=0})^2-({\rm Im}
\mathcal{M}_\theta)^2]_{ii}+{\cal O}(\theta^3)$.
The first 4 terms can be absorbed into the
redefinition of  kinetic and mass terms, as usual 
in the case with $\theta=0$.
The 5th and 6th terms represent a peculiar contribution
due to nonzero $\theta$.
We have kept the constant (but $\theta$ dependent)
terms in the last 3 terms for the calculation of 
topological susceptibility,
which we will address later.

%%%%%%%%%%%%%%%%%%%%%%%%%%%%%%%%%%%%%%%%%%%%%%%%
\subsection{Vacuum shift at one-loop }
%%%%%%%%%%%%%%%%%%%%%%%%%%%%%%%%%%%%%%%%%%%%%%%%
Let us start with ${\cal O}(p^5)$ terms, which 
appears only in the nonzero $\theta$ vacuum:
%%%%%%%%%% Eq Lagrangian_5 %%%%%%%%%%%%%%%%%%%%%%%%%%%%%%%%
\begin{eqnarray}
\mathcal{L}_5 &=& \theta \left[\frac{\sqrt{2}}{6F}\bar{M}^2
{\rm Tr}\xi^3-\frac{\sqrt{2}}{2F}\bar{M}^2
\sum^{N_f}_i (16(N_f L_7+ L_8) M^2_{ii}(\theta)\xi_{ii})\right].
\end{eqnarray}
%where $\bar{M}^2(\theta)\equiv 2\bar{m}(\theta)\Sigma/F^2$.
Taking contractions of $\xi$'s in the first term, 
%it becomes
%%%%%%%%%% Eq Lagrangian_5 contracted %%%%%%%%%%%%%%%%%
%\begin{eqnarray}
%\mathcal{L}_5 &\to& \frac{\sqrt{2}}{2F}\bar{M}^2(\theta)
%\sum^{N_f}_i \left(\langle\xi^2_{ii}\rangle_{\xi}
%-16(N_f L_7+ L_8)M^2_{ii}(\theta)\right)\xi_{ii}
%\nonumber\\
%&=& \frac{\sqrt{2}}{2F}\bar{M}^2(\theta)
%\sum^{N_f}_i \left(\sum^{N_f}_f \Delta(0,M^2_{if}(\theta))
%-G(0,M^2_{ii}(\theta),M^2_{ii}(\theta))
%\right.\nonumber\\&&\left.
%-16(N_f L_7+ L_8)M^2_{ii}(\theta)\right)\xi_{ii}.\nonumber\\
%\end{eqnarray}
and noting $\sum_i\xi_{ii}=0$, it becomes
%%%%%%%%%% Eq Lagrangian_5 rewritten %%%%%%%%%%%%%%%%%
\begin{eqnarray}
\label{eq:L_5eff}
\mathcal{L}_5 &\to& \theta\left[
\sum_i \frac{F}{\sqrt{2}} M^2_{ii}B_{ii}\xi_{ii}(x)
\right]+{\cal O}(\theta^3),
\end{eqnarray}
where
%%%%%%%%%% Eq B_ii definition %%%%%%%%%%%%%%%%%%%%%%%%
\begin{eqnarray}
B_{ii} &\equiv&  
\frac{\bar{m}}{m_i}
\times \frac{1}{F^2}
\left[
  \left(\sum^{N_f}_f \Delta(0,M^2_{if})
-G(0,M^2_{ii},M^2_{ii})
%\right.\right.\nonumber\\&&\left.\left.
-16(N_f L_7+ L_8)M^2_{ii}\right)
\right.\nonumber\\&&\left.
-\frac{1}{N_f}\sum^{N_f}_j  \left(\sum^{N_f}_f \Delta(0,M^2_{jf})
-G(0,M^2_{jj},M^2_{jj})
%\right.\right.\nonumber\\&&\left.\left.
-16(N_f L_7+ L_8)M^2_{jj}\right)\right],\nonumber\\
\label{eq:Bii}
\end{eqnarray}
which gives an ${\cal O}(p^2)$ contribution.
Here $\theta$ dependence of the masses is dropped, 
since it gives only ${\cal O}(\theta^3)\times {\cal O}(p^2)$ contributions;
we have simply set $M^2_{ij}=M^2_{ij}(\theta=0)$ in (\ref{eq:Bii}).
Note also  that $\sum_i M_{ii}^2B_{ii}=0$.\\

The above linear term in $\xi$ in Eq.(\ref{eq:L_5eff}) requires 
further shift in the vacuum $U_0$ ($\times e^{-i\theta/N_f}$); 
%%%%%%%%%% Eq U_0 shift %%%%%%%%%%%%%%%%%%%%%%%%%%%%%%
\begin{eqnarray}
U_0 e^{-i\theta/N_f} \to U_0^\prime \equiv
{\rm diag} (e^{-i\theta_1}, e^{-i\theta_2},\cdots),
\end{eqnarray}
where the phase of $i$th diagonal component
is given by
%%%%%%%%%% Eq theta_i %%%%%%%%%%%%%%%%%%%%%%%%%%%%%%%%
\begin{eqnarray}
\theta_i &\equiv& \frac{\bar{m}}{m_i}\theta -a_i \theta^3
-b_i \theta,
\end{eqnarray}
with
%%%%%%%%% Eq b_i definition %%%%%%%%%%%%%%%%%%%%%%%%%%
\begin{eqnarray}
\label{eq:b_i}
b_i \equiv -B_{ii}+\frac{\bar{m}}{m_i}\sum^{N_f}_f B_{ff}.
\end{eqnarray}
Again, we have ignored ${\cal O}(\theta^3)\times {\cal O}(p^2)$ 
contributions here.

The meson mass is also shifted as
%%%%%%%%%% Eq mass shift %%%%%%%%%%%%%%%%%%%%%%%%%%%%%%
\begin{eqnarray}
M^2_{ij}(\theta) \to (M^\prime_{ij}(\theta))^2 \equiv 
\frac{\Sigma}{F^2}(m_i\cos \theta_i +m_j \cos \theta_j),
\end{eqnarray}
and the effective Lagrangian up to ${\cal O}(p^5)$ then reads
%%%%%%%%%% Eq Lagrangian_5 effective %%%%%%%%%%%%%%%%%%
\begin{eqnarray}
\mathcal{L}_{LO}+\mathcal{L}_5 &=& 
 \frac{1}{2}{\rm Tr}(\partial_\mu \xi(x))^2
+\frac{1}{2}\sum_i^{N_f}(M^{\prime}_{ii}(\theta))^2
%({\rm Re}\mathcal{M^\prime}_\theta)_{ii}
[\xi^2(x)]_{ii} 
\nonumber\\&&
+\frac{\sqrt{2}}{6F}\bar{M}^2\theta
\left({\rm Tr}\xi^3-3\sum^{N_f}_i\langle \xi^2_{ii}\rangle_\xi \xi_{ii}\right).
%\nonumber\\
\end{eqnarray}
\subsection{Inserting sources}
%%%%%%%%%%%%%%%%%%%%%%%%%%%%%%%%%%%%%%%%%%%%%%%%

Next, we consider insertions of the pseudoscalar 
and axial vector sources, $p(x)$ and $a_\mu(x)$.
Since the parity symmetry is broken by the $\theta$
term, we will see that these source terms have
unusual contributions which look 
like scalar or vector operators.
It is therefore convenient to define 
the {\it shifted} Hermitian sources as
%%%%%%% Eq shifted sources %%%%%%%%%%%%%%%%%%%%%
\begin{eqnarray}
p^+(x) &\equiv & \frac{1}{2}(U_0^{\prime \dagger} p(x) 
+ p(x)U^\prime_0), \\
p^-(x) &\equiv & \frac{i}{2}(U_0^{\prime \dagger} p(x) 
- p(x)U^\prime_0), \\
a_\mu^+(x) &\equiv & \frac{1}{2}(U_0^{\prime \dagger} 
a_\mu(x)U^\prime_0 
+ a_\mu(x)), \\
a_\mu^-(x) &\equiv & \frac{i}{2}(U_0^{\prime \dagger} a_\mu(x) 
U^\prime_0- a_\mu(x)), 
\end{eqnarray}
where we have assumed the original $p(x)$ and $a_\mu(x)$ are
both Hermitian and traceless matrices.
In the following, we consider only charged meson 
type sources which have two 
different flavor indices. For this case the absence 
of the diagonal parts:
$[p]_{ii}=[p^+]_{ii}=[p^-]_{ii}=0$
and $[a_\mu]_{ii}=[a^+_\mu]_{ii}=[a_\mu^-]_{ii}=0$ (for all $i$)
simplifies the calculation.\\

\if0

In the same way as the calculation of the Lagrangian
above, the source terms at 1-loop order can be obtained
%%%%%%%% Eq L_src %%%%%%%%%%%%%%%%%%%%%%%%%%%%%%
\begin{eqnarray}
\mathcal{L}_{src}(p,a_\mu)&=& -\frac{\sqrt{2}\Sigma}{F}
\sum_{i,j}^{N_f}\xi_{ij}(x)p^+_{ji}(x)
\nonumber\\&&\times\left[1-\frac{2}{3F^2}
\left(\frac{\sum_f^{N_f}(\Delta(0,M_{if}^2(\theta))
+\Delta(0,M_{jf}^2(\theta)))}{2}
\right.\right.\nonumber\\&&\left.\left.
\hspace{1in}+A(0,M_{ii}^2(\theta),M_{jj}^2(\theta))\right)
\right.\nonumber\\&&\left.
\hspace{-0.5in}+\frac{1}{F^2}\left(G(0,M_{ii}^2(\theta),M_{jj}^2(\theta))
+16(L_6 \sum_f M_{ff}^2(\theta)+L_8M_{ij}^2(\theta))\right)
\right]
\nonumber\\&&
\hspace{-0.5in}+\Sigma{\rm Tr}\left[p^-(x)\left\{\frac{\xi^2(x)}{F^2}
-\frac{\sqrt{2}}{F}\xi(x)
\left(\frac{16(N_fL_7+L_8)}{F^2}\bar{M}^2(\theta)\right)\right\}\right]
\nonumber\\&&
%-\frac{8L_8\Sigma}{F^2}\sum_i [p^-(x)]_{ii}M^2_{ii}(\theta)
%-\frac{\sqrt{2}\Sigma}{F}{\rm Tr}[p^-(x)\xi(x)]
-\sqrt{2}F\sum^{N_f}_{i,j}[\partial_\mu \xi(x)]_{ij}[a^+_\mu(x)]_{ji}
\nonumber\\&&
\times\left[1-\frac{2}{3F^2}
\left(\frac{\sum_f^{N_f}(\Delta(0,M_{if}^2(\theta))
+\Delta(0,M_{jf}^2(\theta)))}{2}
\right.\right.\nonumber\\&&\left.\left.
%\hspace{1in}
+A(0,M_{ii}^2(\theta),M_{jj}^2(\theta))\right)
%\right.\nonumber\\&&\left.
%\hspace{-0.5in}
+\frac{8}{F^2}\left(L_4 \sum_f M_{ff}^2(\theta)+L_5M_{ij}^2(\theta))\right)
\right]
\nonumber\\&&
+\sum^{N_f}_{i,j}[\partial_\mu \xi(x) \xi(x)-\xi(x)\partial_\mu \xi(x)]_{ij}
[a^-_\mu(x)]_{ji}
\end{eqnarray}
\fi

%%%%%%%%%%%%%%%%%%%%%%%%%%%%%%%%%%%%%%%%%%%%%%%%
\subsection{One-loop effective Lagrangian with sources}
%%%%%%%%%%%%%%%%%%%%%%%%%%%%%%%%%%%%%%%%%%%%%%%%

In the expansion of the leading Lagrangian Eq.(\ref{eq:Lagrangian}),
we also have ${\cal O}(p^6)$ terms,
%%%%%%%% Eq Sigma and F NLO %%%%%%%%%%%%%%%%%%%%
\begin{eqnarray}
\frac{1}{6F^2}{\rm Tr}[\partial_\mu\xi\xi\partial_\mu \xi\xi-
\xi^2(\partial_\mu \xi)^2]-\frac{1}{12F^2}
\sum^{N_f}_{i}M^2_{ii}(\theta)[\xi^4]_{ii}.
\end{eqnarray}
from which contribution can be calculated
in a straightforward way as in the case
at $\theta=0$ \cite{Gasser:1983yg}.

Collecting all contributions so far,  the LO + NLO  effective 
Lagrangian, including  the pseudoscalar and axial vector
sources, is given by
%%%%%%%% Eq one-loop L %%%%%%%%%%%%%%%%%%%%
\begin{eqnarray}
\label{eq:effL}
\mathcal{L}^\theta_{eff}(p,a_\mu) &=&
 \frac{1}{2}\sum^{N_f}_{i,j} (Z_\xi^{ij}(\theta))^2
\left([\partial_\mu \xi_{ij}\partial_\mu \xi_{ji}](x)
+(M^{\prime}_{ij}(\theta))^2(Z_M^{ij}(\theta))^2
[\xi_{ij}\xi_{ji}](x)\right) 
\nonumber\\&&
+\frac{\sqrt{2}}{6F}\bar{M}^2\theta
\left({\rm Tr}\xi^3(x)-3\sum^{N_f}_i\langle \xi^2_{ii}\rangle_\xi 
\xi_{ii}(x)\right)
\nonumber\\&&
+\frac{1}{2}\sum^{N_f}_{i,j}[\xi_{ij}\xi_{ji}(x)]\times
\left(-\frac{16}{F^2}(N_fL_7+L_8)\bar{M}^4\theta^2\right) 
\nonumber\\&&
-\frac{\sqrt{2}\Sigma}{F}\sum^{N_f}_{i,j}[p^+_{ji}(x)\xi_{ij}(x)]\times 
Z_\xi^{ij}(\theta)Z_F^{ij}(\theta)(Z_M^{ij}(\theta))^2
\nonumber\\&&
%\hspace{-0.2in}
+\sum^{N_f}_{i,j}\left[p^-_{ji}(x)\left(\frac{\Sigma}{F^2}\xi^2_{ij}(x)
-\frac{16\sqrt{2}\Sigma(N_fL_7+L_8)}{F^3}\bar{M}^2\theta\xi_{ij}(x)\right)
\right]
\nonumber\\&&
-\sqrt{2}F\sum^{N_f}_{i,j}[a^+_\mu(x)_{ji}\partial_\mu \xi_{ij}(x)]\times 
Z_\xi^{ij}(\theta)Z_F^{ij}(\theta)
\nonumber\\&&
+\sum^{N_f}_{i,j}[a^-_\mu(x)]_{ji}[\partial_\mu 
\xi(x) \xi(x)-\xi(x)\partial_\mu \xi(x)]_{ij}
\nonumber\\&&
%\hspace{-0.8in}
+ \frac{1}{2F^2}\sum^{N_f}_{i,j} 
\left([\partial_\mu \xi_{ii}\partial_\mu \xi_{jj}](x)
\frac{\Delta(0,M_{ij}^2(\theta))}{3}
\right.\nonumber\\&&\left.
-\left(\frac{2}{3}M_{ij}^2(\theta)\Delta(0,M_{ij}^2(\theta))
-16L_7M^2_{ii}(\theta)M^2_{jj}(\theta)\right)
[\xi_{ii}\xi_{jj}](x)\right)
\nonumber\\&&
-\Sigma \left(\sum_f^{N_f}m_f \cos \theta_f \right)
-4L_6 (\sum_fM^2_{ff}(\theta))^2
%\nonumber\\&&
%-2L_8\sum_fM^4_{ff}(\theta=0)
+4N_f(N_fL_7+L_8)\bar{M}^4\theta^2,
\end{eqnarray}
where we have omitted $\theta$-independent constants. 
We also have omitted multi $n$ point vertices for $n> 3$,
which are irrelevant for 
the two-point correlation functions below.

In the above result,
the $Z$ factors are given by
%%%%%%%% Z's definition %%%%%%%%%%%%%%%%%%%%%%%%%%%
\begin{eqnarray}
%%% Z xi
Z^{ij}_\xi(\theta) &\equiv& 1-\frac{1}{2F^2}\left[
\frac{\sum_f^{N_f}(\Delta(0,M_{if}^2(\theta))
+\Delta(0,M_{jf}^2(\theta)))}{6}+
\frac{A(0,M_{ii}^2(\theta),M_{jj}^2(\theta))}{3}
\right.\nonumber\\&&\left.
\hspace{1in}-8\left(L_4\sum_f^{N_f}M^2_{ff}(\theta)
+L_5M_{ij}^2(\theta)\right)\right],\\
%%% Z_M
Z^{ij}_M(\theta) &\equiv& 1+\frac{1}{2F^2}\left[
G(0,M^2_{ii}(\theta),M^2_{jj}(\theta))
-8(L_4-2L_6)\sum_f^{N_f}M_{ff}^2(\theta)
%\right.\nonumber\\&&\left.
%\hspace{1in}
-8(L_5-2L_8)M_{ij}^2(\theta)
\right],\nonumber\\\\
%%% Z_F
Z^{ij}_F(\theta) &\equiv& 1-\frac{1}{2F^2}\left[
\frac{\sum_f^{N_f}(\Delta(0,M_{if}^2(\theta))
+\Delta(0,M_{jf}^2(\theta)))}{2}+
A(0,M_{ii}^2(\theta),M_{jj}^2(\theta))
\right.\nonumber\\&&\left.
\hspace{1in}-8\left(L_4\sum_f^{N_f}M^2_{ff}(\theta)
+L_5M_{ij}^2(\theta)\right)\right],
\end{eqnarray}
where
%%%%%%%% Eq A ,A definition %%%%%%%%%%%%%%%%%%%%%
\begin{eqnarray}
A(x, M^2_{ii},M^2_{jj})
&\equiv& G(x,M^2_{ii},M^2_{jj})
%\nonumber\\&&
-\frac{G(x,M^2_{ii},M^2_{ii})
+G(x,M^2_{jj},M^2_{jj})
}{2}.
%\nonumber\\
\end{eqnarray}
and its derivative
%%%%%%% A derivative %%%%%%%%%%%%%%%%%%%%%%%%%%%%%%%%
\begin{eqnarray}
\partial^2_\mu A(x, M^2_{ii},M^2_{j j})
&=& M^2_{ij}G(x,M^2_{ii},M^2_{j j})
\nonumber\\
&&-\frac{
M^2_{ii}G(x,M^2_{ii},M^2_{ii})
+M^2_{j j}G(x,M^2_{j j},M^2_{j j})
}{2},
%\nonumber\\
\end{eqnarray}
are UV finite at $x=0$, 
and both vanish when $M^2_{j j}=M^2_{ii}$.

On the other hand,  $\Delta(0,M^2)$ and $G(0,M^2_1,M^2_2)$
are logarithmically divergent, 
and the divergent parts are evaluated by the dimensional regularization
at $D=4-2\epsilon$ as
%%%%%%%%% Eq Delta, G divergence %%%%%%%%%%%
\begin{eqnarray}
\Delta(0,M^2) &=& -\frac{M^2}{16\pi^2}\left(\frac{1}{\epsilon}+1-\gamma+\ln 4\pi\right) +\cdots,\nonumber\\
G(0,M_1^2,M_2^2) &=& -\frac{1}{16\pi^2}\left(\frac{M_1^2+M_2^2}{N_f}-
\frac{1}{N_f^2}\sum_f^{N_f}M^2_{ff}(\theta)\right)
\left(\frac{1}{\epsilon}+1-\gamma+\ln 4\pi\right)
+\cdots,
\end{eqnarray}
where $\gamma=0.57721\cdots$ denotes Euler's constant.
As is the usual case, these divergence can be removed by
the renormalization of $L_i$'s as
%%%%%% Eq renormalization of L_'s %%%%%%%%%%
\begin{eqnarray}
\label{eq:Lr}
L_i &=& L_i^r(\mu_{sub})-\frac{\gamma_i}{32\pi^2}
\left(\frac{1}{\epsilon}+1-\gamma+\ln 4\pi-\ln \mu^2_{sub}\right),
%L_4 &=& \frac{1}{(16\pi)^2}\ln \Lambda^2 + L_4^r,\nonumber\\
%L_5 &=& \frac{N_f}{(16\pi)^2}\ln \Lambda^2 + L_5^r,\nonumber\\
%L_6 &=& \frac{1}{(16\pi)^2}\left(\frac{1}{2}+\frac{1}{N_f^2}\right)
%\ln \Lambda^2 + L_6^r,\nonumber\\
%L_7 &=& L_7^r,\nonumber\\
%L_8 &=& \frac{1}{(16\pi)^2}\left(\frac{N_f}{2}-\frac{2}{N_f}\right)
%\ln \Lambda^2 + L_8^r,
\end{eqnarray}
where $L^r_i(\mu_{sub})$'s denote the renormalized low energy constants
at the subtraction scale $\mu_{sub}$ and
%%%%%% Eq gamma_i 's  %%%%%%%%%%%%%%%%%%%%%%%%%
\begin{eqnarray}
\gamma_4 = \frac{1}{8}, \;\;
\gamma_5=\frac{N_f}{8},\;\;
 \gamma_6 = \frac{1}{8}\left(\frac{1}{2}+\frac{1}{N_f^2}\right),\;\;
\gamma_7=0, \;\;
\gamma_8=\frac{1}{8}\left(\frac{N_f}{2}-\frac{2}{N_f}\right).
\end{eqnarray}
As a result,
$Z^{ij}_F(\theta)$, $Z^{ij}_M(\theta)$, and $M^\prime_{ij}(\theta)$ are kept finite,
while
$Z^{ij}_\xi(\theta)$ is still divergent but it does not affect 
the physical observables.

After this procedure, one can replace $\Delta(0,M^2)$ by
\begin{eqnarray}
\label{eq:Deltaren}
\Delta^r(0,M^2,\mu^2_{sub}) &=& \frac{M^2}{16\pi^2}\ln\frac{M^2}{\mu_{sub}^2}
+g_1(M^2),
\end{eqnarray}
where $g_1$ denotes the finite volume contribution \cite{Bernard:2001yj}:
%%%%% Eq. g_1 %%%%%%%%%%%%%%%%%%%%%%%%%%%%%%%%%
\begin{eqnarray}
\label{eq:g1p}
g_1(M^2) = \sum_{a\neq 0}\int 
\frac{d^4 q}{(2\pi)^4}\frac{e^{-iqa}}{q^2+M^2}
=\sum_{a\neq 0} \frac{M}{4\pi^2|a|}K_1(M|a|),
\end{eqnarray}
where $K_1$ is the modified Bessel function and
the summation is taken over the four-vector 
$a_\mu = n_\mu L_\mu$ with $L_i = L\; (i=1,2,3)$ and $L_4=T$.
Numerically, truncation at $|n_\mu|\leq 5$
already gives a good accuracy when $ML >3$.
For the explicit expression of $G(0,M_1^2,M_2^2)$,
see Appendix \ref{app:G-property}.

It is noted here that the 8th term of Eq.(\ref{eq:effL}) does not 
contribute to the NLO two-point functions, 
since we consider only off-diagonal sources,
$p(x)_{i\not= j}$ or $a_\mu(x)_{i\not= j}$, 
whose coupling to the diagonal element $\xi_{ii}$ 
is of higher order [next-to-next-to-leading order (NNLO)].

%% file: sec4.tex
%%%%%%%%%%%%%%%%%%%%%%%%%%%%%%%%%%%%%%%%%%%%%%%%%%%
%%%%%%%%%%%%%%%%%%%%%%%%%%%%%%%%%%%%%%%%%%%%%%%%%%%
%%%%%%%%%%%%%%%%%%%%%%%%%%%%%%%%%%%%%%%%%%%%%%%%%%%
\section{Two-point functions}
\label{sec:2pt}
%\setcounter{equation}{0}
%%%%%%%%%%%%%%%%%%%%%%%%%%%%%%%%%%%%%%%%%%%%%%%%%%%a
%%%%%%%%%%%%%%%%%%%%%%%%%%%%%%%%%%%%%%%%%%%%%%%%%%%
%%%%%%%%%%%%%%%%%%%%%%%%%%%%%%%%%%%%%%%%%%%%%%%%%%%

Pseudoscalar and axial vector correlators are obtained by
the functional derivatives of the partition function
%%%%%% Z_src %%%%%%%%%%%%%%%%%%%%%%%%%%%%%%%
\begin{eqnarray}
\mathcal{Z}^\theta(p,a_\mu) &\equiv& 
\int d\xi \;e^{-\int d^4 x \mathcal{L}^\theta_{eff}(p,a_\mu)},
\end{eqnarray}
with respect to the sources $p$ and $a_\mu$.

Here we derive the two-point correlation functions 
of these operators in an irreducible representation
which consist of two different valence quarks. 
We consider the most general nondegenerate case
where their masses are denoted by 
$m_v$ and $m_{v^\prime}$, respectively.
More explicitly, we calculate the three types of
correlation functions with zero momentum projection
(three-dimensional integral),
%%%%%% Eq PP AA definitions %%%%%%%%%%
\begin{eqnarray}
\mathcal{PP}(t, m_v, m_{v^\prime}) &\equiv &
\nonumber\\&&\hspace{-1in}
\left.\frac{1}{2} \int d^3 x \left(\frac{\delta}{\delta p(x)_{v v^\prime}}
+\frac{\delta}{\delta p(x)_{v^\prime v}}\right)
\left(\frac{\delta}{\delta p(0)_{v v^\prime}}
+\frac{\delta}{\delta p(0)_{v^\prime v}}\right)
\ln \mathcal{Z}^\theta(p,a_\mu)\right|_{p=0,a_\mu=0},\\%\nonumber\\\\
\mathcal{A}_0\mathcal{P}(t, m_v, m_{v^\prime}) &\equiv &
\nonumber\\&&\hspace{-1in}
\left.\frac{1}{2} \int d^3 x 
\left(\frac{\delta}{\delta a_0(x)_{v v^\prime}}
+\frac{\delta}{\delta a_0(x)_{v^\prime v}}\right)
\left(\frac{\delta}{\delta p(x)_{v v^\prime}}
+\frac{\delta}{\delta p(0)_{v^\prime v}}\right)
\ln \mathcal{Z}^\theta(p,a_\mu)\right|_{p=0,a_\mu=0},\\ %\nonumber\\\\
\mathcal{A}_0\mathcal{A}_0(t, m_v, m_{v^\prime}) &\equiv &
\nonumber\\&&\hspace{-1in}
\left.\frac{1}{2} \int d^3 x 
\left(\frac{\delta}{\delta a_0(x)_{v v^\prime}}
+\frac{\delta}{\delta a_0(x)_{v^\prime v}}\right)
\left(\frac{\delta}{\delta a_0(0)_{v v^\prime}}
+\frac{\delta}{\delta a_0(0)_{v^\prime v}}\right)
\ln \mathcal{Z}^\theta(p,a_\mu)\right|_{p=0,a_\mu=0}, %\nonumber\\
\end{eqnarray}
where we denote $t=x^0$.
As mentioned before,the partial quenching is performed by the replica trick:
extending the number of flavors $N_f\to N_f+N_v+(N-N_v)$
then taking the limit $N\to 0$.

Noting 
%%%%%%% Eq p and a derivative in theta %%%%%%%%%%%%%%%%
\begin{eqnarray}
\frac{\delta}{\delta p(x)_{vv^\prime }} &=& 
\left(\frac{e^{i\theta_v}+e^{-i\theta_{v^\prime}}}{2}\right)
\frac{\delta}{\delta p^+(x)_{vv^\prime}}
+i\left(\frac{e^{i\theta_v}-e^{-i\theta_{v^\prime}}}{2}\right)
\frac{\delta}{\delta p^-(x)_{vv^\prime}},\\
\frac{\delta}{\delta a_\mu(x)_{vv^\prime }} &=& 
\left(\frac{e^{i(\theta_v-\theta_{v^\prime})}+1}{2}\right)
\frac{\delta}{\delta a_\mu^+(x)_{vv^\prime}}
+i\left(\frac{e^{i(\theta_v-\theta_{v^\prime})}-1}{2}\right)
\frac{\delta}{\delta a_\mu^-(x)_{vv^\prime}},
\end{eqnarray}
the correlation functions are given by 
%(see the appendix \ref{app:Corr} for the details)
%%%%%% Eq PP correlators %%%%%%%%%%%%%%%%%
\begin{eqnarray}
\label{eq:PP}
\mathcal{PP}(t, m_v, m_{v^\prime}) &=&
%\nonumber\\&&\hspace{-1.3in}
%\frac{(\Sigma^{\rm 1-loop}_{vv^\prime}(\theta))^2}{(F Z_F^{vv^\prime}(\theta))^2}
%Z_{PP}^{vv^\prime}(\theta)
[C^{\theta}_{PP}(m_v,m_{v^\prime})]^{\rm 1-loop}
\frac{\cosh(M^{\rm 1-loop}_{vv^\prime}(\theta)(t-T/2))}
{M^{\rm 1-loop}_{vv^\prime}(\theta)
\sinh(M^{\rm 1-loop}_{vv^\prime}(\theta)T/2)}
\nonumber\\&&
+\frac{(\theta_v+\theta_{v^\prime})^2}{4}\frac{\Sigma^2}{F^4}C^{vv^\prime}_X(t),
%\nonumber\\&&\hspace{-.4in}
%-(\theta_v+\theta_{v^\prime})\frac{\Sigma^2\bar{M}^2(\theta)}{F^4}
%C_{\Delta X}^{vv^\prime}(t,M_{vv^\prime}^2)
%+\frac{\Sigma^2\bar{M}^4(\theta)}{F^4}(-\partial_{M^2})
%C_{\Delta X}^{vv^\prime}(t,M^2)|_{M=M_{vv^\prime}},
%\nonumber\\
\\
%\end{eqnarray}
%%%%%%% Eq AP correlators %%%%%%%%%%%%%%%%%
%\begin{eqnarray}
\label{eq:AP}
\mathcal{A}_0\mathcal{P}(t, m_v, m_{v^\prime}) %&& %
&=&
%\nonumber\\&&\hspace{-1.3in}=
%\Sigma^{\rm 1-loop}_{vv^\prime}(\theta)Z_{AP}^{vv^\prime}(\theta)
[C^{\theta}_{AP}(m_v,m_{v^\prime})]^{\rm 1-loop}
\frac{\sinh(M^{\rm 1-loop}_{vv^\prime}(\theta)(t-T/2))}
{\sinh(M^{\rm 1-loop}_{vv^\prime}(\theta)T/2)}
%-\frac{(\theta_v+\theta_{v^\prime})}{2}\frac{\Sigma\bar{M}^2(\theta)}
%{F^2}\partial_tC^{vv^\prime}_{\Delta X}(t,M_{vv^\prime}^2)
%\nonumber\\&&\hspace{-1.4in}
%-\frac{\Sigma\bar{M}^4(\theta)}{2M_{vv^\prime}}\frac{\partial}{\partial M}
%\left.\partial_tC^{vv^\prime}_{\Delta X}(t,M^2)\right|_{M=M_{vv^\prime}}
\nonumber\\&&
-\frac{(\theta_v^2-\theta_{v^\prime}^2)}{4}\frac{\Sigma^2}{F^2}
C_{Y_0}^{vv^\prime}(t),\\
%-\frac{(\theta_v-\theta_{v^\prime})}{2}\frac{\Sigma^2\bar{M}^2(\theta)}{F^2}
%C_{Y_0\Delta}^{vv^\prime}(t),
%\nonumber\\
%\end{eqnarray}
%and
%%%%%% Eq AA correlators %%%%%%%%%%%%%%%%%
%\begin{eqnarray}
\label{eq:AA}
\mathcal{A}_0\mathcal{A}_0(t, m_v, m_{v^\prime}) %&&
&=&
%\nonumber\\&&\hspace{-1.3in}=
%-(F Z_F^{vv^\prime}(\theta))^2Z_{AA}^{vv^\prime}(\theta)
[C^{\theta}_{AA}(m_v,m_{v^\prime})]^{\rm 1-loop}
M^{\rm 1-loop}_{vv^\prime}(\theta)
\frac{\cosh(M^{\rm 1-loop}_{vv^\prime}(\theta)(t-T/2))}
{\sinh(M^{\rm 1-loop}_{vv^\prime}(\theta)T/2)}
%-(\theta_v-\theta_{v^\prime})\bar{M}^2(\theta)
%\partial_tC^{vv^\prime}_{Y_0\Delta}(t)
%\nonumber\\&&
%\hspace{-1.4in}
%+\frac{\Sigma\bar{M}^4(\theta)}{2M_{vv^\prime}}\frac{\partial}{\partial M}
%\left.\partial^2_tC^{vv^\prime}_{\Delta X}(t,M^2)\right|_{M=M_{vv^\prime}}
\nonumber\\&&
-\frac{(\theta_v-\theta_{v^\prime})^2}{4}
C_{W_{00}}^{vv^\prime}(t).
%\nonumber\\
\end{eqnarray}
Here, $[C^{\theta}_{JJ^\prime}(m_v,m_{v^\prime})]^{\rm 1-loop}$'s denote
the overall coefficients of which definitions are given in 
Appendix~\ref{app:Corr}.

Because of the {\it CP} violation, each correlator has a contribution from the
two-pion state's propagation denoted by 
$C^{vv^\prime}_X(t)$, $C^{vv^\prime}_{Y_0}(t)$ 
and $C_{W_{00}}^{vv^\prime}$ respectively, 
which are of ${\cal O}(e^{-2Mt})$ at large time separation
(See Appendix~\ref{app:Corr} for the details.).

The correction to the pseudo--Nambu-Goldstone 
boson mass is given by
\begin{eqnarray}
M^{\rm 1-loop}_{vv^\prime}(\theta) &\equiv & 
M^\prime_{vv^\prime}(\theta)%\times 
Z_M^{vv^\prime}(\theta) %\times 
\left(1-\frac{\bar{M}^4\theta^2\left\{32(N_fL^r_7(\mu_{sub})+L^r_8(\mu_{sub}))
+H^r_{vv^\prime}(M^2_{vv^\prime},\mu_{sub})\right\}}
{4F^2M^2_{vv^\prime}}
\right),\nonumber\\
\end{eqnarray}
and the pion decay constant is extracted 
from the $\mathcal{PP}(t, m_v, m_{v^\prime})$ and 
$\mathcal{AP}(t, m_v, m_{v^\prime})$ correlators in a standard way as
\begin{eqnarray}
F^{\rm 1-loop}(\theta) &\equiv& F Z_F^{vv^\prime}(\theta) \times 
\frac{Z^{vv^\prime}_{AP}(\theta)}{\sqrt{Z_{PP}^{vv^\prime}(\theta)}}\nonumber\\
&=&  F Z_F^{vv^\prime}(\theta) \left[1-\frac{1}{8}(\theta_v-\theta_{v^\prime})^2
-\frac{\bar{M}^4\theta^2}{4F^2}\left(\left.\frac{\partial}{\partial M^2}
H^r_{vv^\prime}(M^2,\mu_{sub})\right|_{M=M_{vv^\prime}}\right)
\right.\nonumber\\&&\left.
+\frac{1}{4}(\theta_v-\theta_{v^\prime})\frac{\bar{M}^2\theta}{F^2}
H^\prime_{vv^\prime}(M^2_{vv^\prime})
\right].
\end{eqnarray}
The definitions of $H_{ij}(M^2)$ and $H^\prime_{ij}(M^2)$
are given in Appendix~\ref{app:Corr}.

One should note that for very small $\theta$, 
the above formulas can be greatly simplified by
ignoring ${\cal O}(\theta^2)\times {\cal O}(p^2)$
corrections or just by setting $\bar{M}^2\theta=0$ and 
$M^\prime_{vv^\prime}(\theta)=M_{vv^\prime}(\theta)$.

%% file: sec5.tex
%%%%%%%%%%%%%%%%%%%%%%%%%%%%%%%%%%%%%%%%%%%%%%%%%%%
%%%%%%%%%%%%%%%%%%%%%%%%%%%%%%%%%%%%%%%%%%%%%%%%%%%
\section{Fixed topology}
\label{sec:fixedQ}
%\setcounter{equation}{0}
%%%%%%%%%%%%%%%%%%%%%%%%%%%%%%%%%%%%%%%%%%%%%%%%%%%
%%%%%%%%%%%%%%%%%%%%%%%%%%%%%%%%%%%%%%%%%%%%%%%%%%%
%%%%%%%%%%%%%%%%%%%%%%%%%%%%%%%%%%%%%%%%%%%%%%%%%%%

From the $\theta$ dependence obtained so far,
we can derive the correlators in a fixed 
topological sector of $Q$.
It is known that an observable at a fixed topology 
(let us denote $G_Q$) 
is related to the one in the $\theta$ vacuum [$G(\theta)$]
by a formula \cite{Aoki:2007ka},
%%%%%%% Eq G_Q general %%%%%%%%%%%
\begin{eqnarray}
\label{eq:G_Qgeneral}
G_Q &=& G(\theta=0)+\frac{\partial^2}{\partial \theta^2} 
G(\theta)|_{\theta=0}\frac{1}{2\chi_t V}
\left[1-\frac{Q^2}{\chi_tV}-\frac{c_4}{2\chi_t^2 V}\right] 
+\frac{\partial^4}{\partial \theta^4} 
G(\theta)|_{\theta=0}\frac{1}{8\chi_t^2V^2} +{\cal O}(V^{-3}),
\nonumber\\
\end{eqnarray}
which is valid in the general theories.
Here, $\chi_t \equiv \langle Q^2 \rangle /V$ denotes
the topological susceptibility and $c_4$ is 
the coefficient of $\theta^4$ term of the 
vacuum energy of the theory.

One should note that in ChPT, the $\theta$ dependence only appears in
the mass term, so that we can treat
$\chi_t \sim c_4 \sim {\cal O}(\mathcal{M})\sim {\cal O}(p^2)$
in the $p$ expansion. Therefore the factor $1/(\chi_t V) $
reduces the order of each  contribution from fixing topology by ${\cal O}(p^2)$,
since $1/V = O(p^4)$ in the $p$ expansion
\footnote{
In the very vicinity of the chiral limit (the $\epsilon$-regime),
$1/ \chi_t V$ becomes ${\cal O}(1)$ and cannot be treated
as perturbation. Exact integrals over $\theta$
is then needed, which is expressed by 
modified Bessel functions 
\cite{e-regime}.
}.
As a consequence, for nonvanishing $G(\theta=0)$,
one can easily calculate the ``NNLO'' contribution 
from fixing topology with the one-loop level calculation only.

%%%%%%%%%%%%%%%%%%%%%%%%%%%%%%%%%
%\subsection{$\theta$ dependence of the vacuum energy}
\subsection{$\chi_t$ at one-loop}
%%%%%%%%%%%%%%%%%%%%%%%%%%%%%%%%%

Let us first calculate $\chi_t$ within
chiral perturbation theory at NLO \cite{Mao:2009sy}.
For $c_4$, as explained above, the tree level calculation
we have given in Eq.(\ref{eq:c4tree}) is enough upto NNLO corrections.
By switching off the source terms, one obtains
%We first calculate 1-loop contributions to $\chi_t$ as
% \begin{eqnarray}
% \chi_t^{\rm 1-loop} &=& - \left. \frac{1}{V}\frac{d^2}{d\theta^2} \log Z^{\rm 1-loop} \right\vert_{\theta = 0}
% =\frac{1}{2}\left. \frac{d^2 }{d\theta^2}M_{ij}^2(\theta)\right\vert_{\theta = 0}
% \langle \xi_{ij}(0)\xi_{ji}(0)\rangle \nonumber \\
% &=&-\bar{m}\Sigma \frac{1}{F^2} 
%\left[\frac{1}{2}\sum^{N_f}_{i,j} 
%\left(\frac{\bar{m}}{m_i}+\frac{\bar{m}}{m_j}\right)
%\Delta(0,M_{ij}^2) -\sum^{N_f}_i 
%\frac{\bar{m}}{m_i}G(0,M^2_{ii},M_{ii}^2)
%\right] .
% \end{eqnarray}
%we calculate the vacuum energy at 1-loop as follows.
%\begin{eqnarray}
%e^{- V_0^{\rm 1-loop}} &=& \int {\cal D}\xi\, \delta({\rm tr}\, \xi )\exp [ - S_{\rm LO} ]
%= \int {\cal D}\xi\, {\cal D} \lambda\, \exp \left[ - S_{\rm LO} + i\int d^4 x \lambda(x)\,
%{\rm tr}\, \xi \right]\nonumber \\
%&=& \int {\cal D}\tilde\xi\, {\cal D} \lambda\, \exp \left[ -\frac{1}{2}\int d^4x\, \left\{\tilde\xi_{ij} (x) D_{ij}(x)
%\tilde \xi_{ji}(x) +\lambda(x) {\rm tr}\, D^{-1}(x-y) \lambda(y)\right\}\right] ,
%\end{eqnarray}
%where $\tilde \xi_{ij} = \xi_{ij} -i\delta_{ij} D_{ii}^{-1}\lambda$. The vacuum energy at 1-loop thus becomes
%\begin{eqnarray}
%V_0^{\rm 1-loop} &=& \frac{1}{2} \sum_{ij}{\rm Tr}\, \log\, D_{ij} + \frac{1}{2}\sum_{ij}{\rm Tr}\, \log\, \sum_f D_{ff}^{-1} . 
%\end{eqnarray}
%from Adding contributions 
from the mass term and terms in the last line 
of Lagrangian Eq.(\ref{eq:effL}),
%to $\chi_t^{\rm 1-loop}$,  
%we obtain the total contribution at NLO as
%%%%%%% Eq chi_t at NLO %%%%%%%%%%%%%%
\begin{eqnarray}
\chi^{\rm 1-loop}_t 
&=& - \left. \frac{1}{V}\frac{d^2}{d\theta^2} 
\ln \mathcal{Z}^\theta(p=0,a_\mu=0) \right\vert_{\theta = 0}
\nonumber\\&=&
-\Sigma \left[\sum^{N_f}_{i}m_i 
\left.\frac{d^2}{d\theta^2}\cos \theta_i\right\vert_{\theta = 0}
\left(1-\frac{\sum^{N_f}_{j}\langle \xi_{ij}(0)
\xi_{ji}(0)\rangle_\xi^{\theta=0}}{F^2}\right)\right]
\nonumber\\
&&-\left.\frac{d^2}{d\theta^2}
\left[4L_6\left(\sum^{N_f}_{i} M_{ii}^2(\theta)\right)^2+
4N_f(N_fL_7+L_8)\bar{M}^4\theta^2\right]\right\vert_{\theta = 0}
\nonumber\\&=&
\bar{m}\Sigma \left[1 -\frac{1}{F^2} 
\left(\frac{1}{2}\sum^{N_f}_{i,j} 
\left(\frac{\bar{m}}{m_i}+\frac{\bar{m}}{m_j}\right)
\Delta(0,M_{ij}^2) -\sum^{N_f}_i 
\frac{\bar{m}}{m_i}G(0,M^2_{ii},M_{ii}^2)
\right.\right.\nonumber\\&&\left.\left.\hspace{1in}
-16L_6 \sum^{N_f}_i M_{ii}^2-16N_f(N_fL_7+L_8)\bar{M}^2
%\left(\frac{2\bar{m}\Sigma}{F^2}\right)
\right)\right], 
\end{eqnarray}
where all the masses in the last line are those at $\theta=0$, 
or namely $M_{ij}^2=M_{ij}^2(\theta=0)$.
In the above calculation, we have used the fact that $\sum_i b_i =0$ [See Eq.(\ref{eq:b_i})].
Note again that the UV divergence is precisely canceled
by the renormalization of $L_i$'s
and therefore, one can replace $\Delta(0,M^2)$
and $L_i$'s with the renormalized values given in
Eqs.(\ref{eq:Deltaren}) and (\ref{eq:Lr}), respectively.
Since the topological susceptibility is
a coefficient of the $\theta^2$ term in the QCD vacuum energy,
it does not depend on $L_4$ and $L_5$ at NLO, which
only appear as corrections to the kinetic term.
%Since the topological susceptibility is 
%a coefficient of the $\theta^2$ term in the QCD vacuum energy,
%it does not depend on $L_4$ and $L_5$ at NLO which 
%only appear as corrections of the kinetic term.

%%%%%%%%%%%%%%%%%%%%%%%%%%%%%%%%%%%%%%%%%%%%%%%%%%%%
\subsection{NLO correction from fixing topology}
%%%%%%%%%%%%%%%%%%%%%%%%%%%%%%%%%%%%%%%%%%%%%%%%%%%%

As discussed above, the next-leading order correction
from fixing topology can be calculated 
at the tree-level.
The above formula is then simplified:
%%%%%%% Eq G_Q general %%%%%%%%%%%
\begin{eqnarray}
\label{eq:fixedQChPT}
G_Q &=& G(\theta=0)+\frac{\partial^2}{\partial \theta^2} 
G(\theta)|_{\theta=0}\frac{1}{2\chi_t V}
\left(1-\frac{Q^2}{\chi_tV}\right) + \mbox{NNLO terms},
\end{eqnarray}
where $\chi_t =\chi_t^{\rm LO}=\bar m \Sigma$ is used.
Furthermore, we can ignore all 
$NLO \times {\cal O}(\theta^2)$ terms
in the correlators $G(\theta)$. 
Note here that we could have omitted $Q^2$ term for small $Q$
but kept it, since it gives 
a $\langle Q^2 \rangle = \chi_t V\sim {\cal O}(1/p^2)$ 
contribution when the topology is summed over again
in the $\theta$ vacuum.

Substituting the expressions in the previous section
into Eq.~(\ref{eq:fixedQChPT}), we obtain the correlators
in a fixed topological sector of $Q$:
%%%%%% Eq PP correlators %%%%%%%%%%%%%%%%%
\begin{eqnarray}
\langle\mathcal{PP}(t, m_v, m_{v^\prime})\rangle_Q &=&
%\frac{(\Sigma^R_{vv^\prime})^2}{(F^R_{vv^\prime})^2}Z_{PP}^{vv^\prime}
C^Q_{PP}(m_v,m_{v^\prime})
\frac{\cosh(M^Q_{vv^\prime}(t-T/2))}{M^Q_{vv^\prime}
\sinh(M^Q_{vv^\prime}T/2)},\\
%\end{eqnarray}
%%%%%% Eq AP correlators %%%%%%%%%%%%%%%%%
%\begin{eqnarray}
\langle\mathcal{A}_0\mathcal{P}(t, m_v, m_{v^\prime})\rangle_Q &=&
C^Q_{AP}(m_v,m_{v^\prime})
\frac{\sinh(M^Q_{vv^\prime}(t-T/2))}
{\sinh(M^Q_{vv^\prime}T/2)},\\
%\end{eqnarray}
%%%%%% Eq AA correlators %%%%%%%%%%%%%%%%%
%\begin{eqnarray}
\langle\mathcal{A}_0\mathcal{A}_0(t, m_v, m_{v^\prime})\rangle_Q &=&
C^Q_{AA}(m_v,m_{v^\prime})
\frac{M^Q_{vv^\prime}\cosh(M^Q_{vv^\prime}(t-T/2))}
{\sinh(M^Q_{vv^\prime}T/2)},
\end{eqnarray}
where the valence pion mass at fixed topology is given by
%%%%%% Eq M^Q %%%%%%%%%%%%%%%%%%%%%%%%%%%%%
\begin{eqnarray}
\label{eq:MpiQ}
(M^Q_{vv^\prime})^2 %&\equiv&
&=& 
(M^{\rm 1-loop}_{vv^\prime}(\theta=0))^2 
%+(\Delta M^{Q}_{vv^\prime})_{\rm NLO}^2,\\
%(M^{\rm 1-loop}_{vv^\prime}(\theta=0))^2&=&
%M_{vv^\prime}^2\left[1+\frac{1}{F^2}\left(
%G(0,M^2_{vv},M^2_{v^\prime v^\prime})
%\right.\right.\nonumber\\&&\left.\left.
%-8(L_4-2L_6)\sum_f^{N_f}M_{ff}^2
%-8(L_5-2L_8)M_{v v^\prime}^2
%\right)\right],\\
%(\Delta M^{Q}_{vv^\prime})_{\rm NLO}^2&=&
\left[1
-%M_{vv^\prime}^2(\theta=0)
\frac{1}{2\chi_t V}
\left(\frac{\bar{m}^2}{m_vm_{v^\prime}}\right)
\left(1-\frac{Q^2}{\chi_t V}\right)
\right],
%\nonumber\\&=&-M_{vv^\prime}^2\left(\frac{\bar{m}^2}{m_vm_{v^\prime}}\right)
%\frac{1}{2\bar{m}\Sigma V}
%\left(1-\frac{Q^2}{\bar{m}\Sigma V}\right),
\end{eqnarray}
and
%%%%%%% C^Q's %%%%%%%%%%%%%%%%%%%%%%%%%%%%%%
\begin{eqnarray}
C^Q_{PP}(m_v,m_{v^\prime})
&\equiv&
%\frac{1}{M^Q_{vv^\prime}}
\left(\frac{\Sigma^{\rm 1-loop}_{vv^\prime}(\theta=0)}
{FZ_F^{vv^\prime}(\theta=0)}
\right)^2
\left[1-\frac{1}{4 \chi_t V}
\left(\frac{\bar m}{m_v}
+\frac{\bar m}{m_{v^\prime}}\right)^2
\left(1-\frac{Q^2}{\chi_t V}\right)\right],\\
%\nonumber\\
%&=& \frac{1}{M^Q_{vv^\prime}}
%\left(\frac{\Sigma^{\rm 1-loop}_{vv^\prime}(\theta=0)}{F^R_{vv^\prime}(\theta=0)}
%\right)^2
%\left[1-\frac{1}{4\bar{m}\Sigma V}
%\left(\frac{1}{m_v}
%+\frac{1}{m_{v^\prime}}\right)^2
%\left(1-\frac{Q^2}{\bar{m}\Sigma V}\right)\right],
%\nonumber\\\\
C^Q_{AP}(m_v,m_{v^\prime})
&=& 
\Sigma^{\rm 1-loop}_{vv^\prime}(\theta=0)
\left[1-\frac{1}{4\chi_t V}
\left(\frac{\bar{m}^2}{m_v^2}
+\frac{\bar{m}^2}{m_{v^\prime}^2}\right)
\left(1-\frac{Q^2}{\chi_t V}\right)\right],\\
C^Q_{AA}(m_v,m_{v^\prime})
&=& 
-(F Z_F^{vv^\prime}(\theta=0))^2%M^Q_{vv^\prime}
\left[1-\frac{1}{4\chi_t V}
\left(\frac{\bar{m}}{m_v}
-\frac{\bar{m}}{m_{v^\prime}}\right)^2
\left(1-\frac{Q^2}{\chi_t V}\right)\right],
\end{eqnarray}
where $\chi_t =\bar{m}\Sigma$.
%Here we omitted the $\theta=0$ dependence of the
%masses $M_{ij}^2=M_{ij}^2(\theta=0)$.
%Note that the first line in Eq.(\ref{eq:MpiQ})
%gives the same NLO mass
%as in the $\theta=0$ vacuum.
%The second line contribution denotes the contribution
%from fixing topology, which, of course, becomes zero
Note that each correction vanishes 
when summed over the topology, since 
$\langle Q^2\rangle = \chi_t V$.

From the above $Q$ dependent correlators,
the conventional extraction of the pion decay 
constant, of course, receives a correction
from fixing the topology:
%%%%%%%% F^Q %%%%%%%%%%%%%%%%%%%%%
\begin{eqnarray}
F^Q_{vv^\prime} &\equiv&
\frac{C^Q_{AP}(m_v,m_{v^\prime})}
{\sqrt{M^Q_{vv^\prime}C^Q_{PP}(m_v,m_{v^\prime})}}
\nonumber\\
&=& F Z_F^{vv^\prime}(\theta=0)
\left[1-\frac{1}{8\chi_t V}
\left(\frac{\bar m}{m_v}
-\frac{\bar m}{m_{v^\prime}}\right)^2
\left(1-\frac{Q^2}{\chi_t V}\right)\right].
\end{eqnarray}
Note that the correction at NLO disappears when $m_v=m_{v^\prime}$.

%% file: sec5-2.tex
%%%%%%%%%%%%%%%%%%%%%%%%%%%%%%%%%%%%%%%%%%%%%%%%%%%
\subsection{NNLO corrections from fixing topology}
%\setcounter{equation}{0}
%%%%%%%%%%%%%%%%%%%%%%%%%%%%%%%%%%%%%%%%%%%%%%%%%%%

In this subsection, we discuss NNLO corrections
from fixing topology to the two-point correlators.
Here we do not calculate two-loop diagrams at $\theta=0$.
They are already known in $N_f=$2 and 2+1 theories
\cite{Bijnens:2004hk}-\cite{Kaiser:2007kf}.
%Here we don't calculate 2-loop diagrams at $\theta=0$, 
%since they are already known 
%\cite{Bijnens:2004hk}-\cite{Kaiser:2007kf}. 
Hereafter we denote them with a superscript ``two-loop.''

Ignoring the multipion states, the functional form
of the correlators at two-loop in the $\theta$ vacuum 
has the following form:
%%%%%% G %%%%%%%%%%%%%%%%%%%%%
\begin{eqnarray}
G(\theta, t) &=& C(\theta)f(M(\theta), t),
\end{eqnarray}
where $C(\theta)$ denotes the time-independent coefficient,
while $f(M(\theta), t)$ represents the one-particle
propagator with a mass $M(\theta)$.

In a fixed topological sector, however,
the correction can not be factorized as
$C^Q \times f(M_Q, t)$ at two-loop level or more.
Using notations
%%%%% del_Q and O^(n) %%%%%%%%%%%%%%%%%%%%%%%%%
\begin{eqnarray}
\delta_Q &\equiv& \frac{1}{2\chi^{\rm 1-loop}_t V}
\left[1-\frac{Q^2}{\chi^{\rm 1-loop}_t V}
-\frac{c_4}{2(\chi^{\rm LO}_t)^2 V}\right],\;\;\;\;\;
%\delta_Q &\equiv& \frac{1}{2\chi_t V}
%\left[1-\frac{Q^2}{\chi_tV}-\frac{c_4}{2\chi_t^2 V}\right],\;\;\;\;\;
O^{(n)}(\theta) \equiv \frac{\partial^n}{\partial \theta^n} O(\theta),
\end{eqnarray}
for an arbitrary function $O$ of $\theta$, and directly substituting the above expression
into Eq.~(\ref{eq:G_Qgeneral}), one obtains at NNLO that
%%%%%% G at 2-loop %%%%%%%%%%%%
\begin{eqnarray}
\label{eq:G2loop}
G_Q(t) &=& C^Q\left.\left[1+D^Q\frac{\partial}{\partial M} 
+\frac{3}{2}(M^{(2)}(0))^2\delta_Q^2\frac{\partial^2}{\partial M^2} \right]
f(M,t)\right|_{M=M(\theta=0)},
\end{eqnarray}
where
%%%%%% C^Q, D_Q, del_Q, 
\begin{eqnarray}
C^Q &\equiv& C(0)+C^{(2)}(0)\delta_Q+C^{(4)}\frac{\delta_Q^2}{2},\\
D^Q &\equiv& M^{(2)}(0)\delta_Q
+\left(M^{(4)}(0)+4M^{(2)}(0)\frac{C^{(2)}(0)}{C(0)}\right)\frac{\delta_Q^2}{2}.
\end{eqnarray}
Note that due to $C^{(2)}(0)/C(0)$ dependence in $D^Q$,  
which is channel dependent, 
and the second derivative term, the correction in Eq.(\ref{eq:G2loop})
cannot be simply absorbed into the mass shift in $f(M,t)$.\\

For more explicit expressions,
we have to calculate the 2nd and 4th derivatives
of various quantities with respect to $\theta$.
We summarize them in Appendix~\ref{app:theta}.
Using the notations there, the correlators 
at NNLO in a fixed topological sector $Q$ are given by
%%%%%% Eq. 2-loop correlators at theta=0 %%%%%%%%%%%
%%%%%% Eq JJ^prime correlators %%%%%%%%%%%%%%%%%
\begin{eqnarray}
\label{eq:corrNNLO}
\langle\mathcal{JJ^\prime}(t, m_v, m_{v^\prime})\rangle^{\rm NNLO}_Q 
&=&
%&&\nonumber\\&&\hspace{-2in}=
%\frac{(\Sigma^R_{vv^\prime})^2}{(F^R_{vv^\prime})^2}Z_{PP}^{vv^\prime}
[C^Q_{JJ^\prime}(m_v,m_{v^\prime})]^{\rm NNLO}
\nonumber\\&&\hspace{-.6in}\times
\left.\left[1+[D^Q_{vv^\prime}]_{JJ^\prime}\frac{\partial}{\partial M} 
+\frac{3}{2}([M_{vv^\prime}^{(2)}]_{LO})^2
\delta_Q^2\frac{\partial^2}{\partial M^2} \right]
f^{JJ^\prime}(M,t)\right|_{M=M^{\rm 2-loop}_{vv^\prime}(\theta=0)},\\{}
[D^Q_{vv^\prime}]_{JJ^\prime} &=&
[M_{vv^\prime}^{(2)}]_{NLO}\delta_Q 
+ \left([M_{vv^\prime}^{(4)}]_{LO}+4[M^{(2)}_{vv^\prime}]_{LO}
[Z_{JJ^\prime}^{vv^\prime}]^{(2)}_{LO}\right)\frac{\delta_Q^2}{2},
\end{eqnarray}
where $J$ and $J^\prime$ represent the operators $P$ or $A$,
%%%%% f^JJ %%%%%%%%%%%%%%%%%%%%%%%%%%%%%%%%%
\begin{eqnarray}
f^{PP}(M,t) &=& \frac{\cosh(M(t-T/2))}{M\sinh(MT/2)},\\
f^{AP}(M,t) &=& \frac{\sinh(M(t-T/2))}{\sinh(MT/2)},\\
f^{AA}(M,t) &=& \frac{M\cosh(M(t-T/2))}{\sinh(MT/2)},
\end{eqnarray}
and
%%%%%%%%% C_JJ^Q NNLO %%%%%%%%%%%%%%%%%%%%%%
\begin{eqnarray}
%%%%%%%%%% C_PP
[C^Q_{PP}(m_v,m_{v^\prime})]^{\rm NNLO}
&=& [C^{\theta =0}_{PP}(m_v,m_{v^\prime})]^{\rm 2-loop}\times
%\left(\frac{M^{\rm 2-loop}_{vv^\prime}
%(\theta=0)}{[M^Q_{vv^\prime}]^{\rm NNLO}_{PP}}\right)
\nonumber\\&&%\hspace{-1.2in}
\left[1+\left([Z_{PP}^{vv^\prime}]^{(2)}_{NLO}
+\frac{4[Z_M^{vv^\prime}]^{(2)}}
{Z_M^{vv^\prime}(\theta=0)}
+\frac{2[Z_F^{vv^\prime}]^{(2)}}
{Z_F^{vv^\prime}(\theta=0)}
\right)\delta_Q
+[Z_{PP}^{vv^\prime}]^{(4)}_{LO}\frac{\delta_Q^2}{2}
\right],\nonumber\\\\{}
%%%%%%%%%% C_AP
[C^Q_{AP}(m_v,m_{v^\prime})]^{\rm NNLO}
&=& [C^{\theta =0}_{AP}(m_v,m_{v^\prime})]^{\rm 2-loop}\times
\nonumber\\&&
\left[1+\left([Z_{AP}^{vv^\prime}]^{(2)}_{NLO}
+\frac{2[Z_M^{vv^\prime}]^{(2)}}
{Z_M^{vv^\prime}(\theta=0)}
+\frac{2[Z_F^{vv^\prime}]^{(2)}}
{Z_F^{vv^\prime}(\theta=0)}
\right)\delta_Q
+[Z_{AP}^{vv^\prime}]^{(4)}_{LO}\frac{\delta_Q^2}{2}
\right],\nonumber\\\\{}
%%%%%%%%%% C_AA
[C^Q_{AA}(m_v,m_{v^\prime})]^{\rm NNLO}
&=& [C^{\theta =0}_{AA}(m_v,m_{v^\prime})]^{\rm 2-loop}\times
\nonumber\\&&
\left[1+\left([Z_{AA}^{vv^\prime}]^{(2)}_{NLO}
+\frac{2[Z_F^{vv^\prime}]^{(2)}}
{Z_F^{vv^\prime}(\theta=0)}
\right)\delta_Q
+[Z_{AA}^{vv^\prime}]^{(4)}_{LO}\frac{\delta_Q^2}{2}
\right].%\nonumber\\
\end{eqnarray}

As is the mass correction, the decay constant 
at fixed topology is not uniquely extracted from the correlators.
If one has a good control of $t$ dependence in
Eq.(\ref{eq:corrNNLO}) on the lattice, however,
a choice is to extract it from the coefficients 
$[C^Q_{PP}(m_v,m_{v^\prime})]^{\rm NNLO}$ and 
$[C^Q_{AP}(m_v,m_{v^\prime})]^{\rm NNLO}$ as
%%%%%%%%%% Eq. F^Q
\begin{eqnarray}
\frac{[C^Q_{AP}(m_v,m_{v^\prime})]^{\rm NNLO}}
{\sqrt{[C^Q_{PP}(m_v,m_{v^\prime})]^{\rm NNLO}}}
&=& F[Z_F^{vv^\prime}(\theta=0)]^{\rm 2-loop}
%\nonumber\\&&\hspace{-1.2in}
\left[
1+\frac{1}{2\chi^{\rm 1-loop}_t V}
\left(1-\frac{Q^2}{\chi_tV}-\frac{c_4}{2\chi_t^2 V}\right)
\right.\nonumber\\&&\left.\times
\left\{[Z_{AP}^{vv^\prime}]^{(2)}_{NLO}-\frac{1}{2}[Z_{PP}^{vv^\prime}]^{(2)}_{NLO}
+\frac{[Z_F^{vv^\prime}]^{(2)}}
{Z_F^{vv^\prime}(\theta=0)}
\right\}
\right.\nonumber\\&&\left.
+\frac{1}{8\chi_t^2 V^2}\left([Z_{AP}^{vv^\prime}]^{(4)}_{LO}
-\frac{1}{2}[Z_{PP}^{vv^\prime}]^{(4)}_{LO}\right)
\right].
\end{eqnarray}

In spite of a complicated channel dependence and
nontrivial $t$ dependence, 
one can see that the (axial) 
Ward-Takahashi identity is kept even at
fixed topology; 
%%%%% Eq WTI %%%%%%%%%%%%%%%%%%%%%%%%%%%%%
\begin{eqnarray}
\label{eq:WTI}
\partial_t \langle\mathcal{A}_0
\mathcal{P}(t, m_v, m_{v^\prime})\rangle_Q &=&
(m_v+m_{v^\prime})
\langle\mathcal{PP}(t, m_v, m_{v^\prime})\rangle_Q,
\end{eqnarray}
which can be easily checked by starting from
time derivative of the correlator in the $\theta$ vacuum, 
%%%%% Eq dAP %%%%%%%%%%%%%%%%%%%%%%%%%%%%%%%%
%\begin{eqnarray}
$
\partial_t \langle\mathcal{A}_0
\mathcal{P}(t, m_v, m_{v^\prime})\rangle_\theta,
$
%\end{eqnarray}
and performing integrals over $\theta$.

%% file: sec6.tex
%%%%%%%%%%%%%%%%%%%%%%%%%%%%%%%%%%%%%%%%%%%%%%%%%%%
%%%%%%%%%%%%%%%%%%%%%%%%%%%%%%%%%%%%%%%%%%%%%%%%%%%
\section{Conclusion}
\label{sec:conclusion}
%\setcounter{equation}{0}
%%%%%%%%%%%%%%%%%%%%%%%%%%%%%%%%%%%%%%%%%%%%%%%%%%%
%%%%%%%%%%%%%%%%%%%%%%%%%%%%%%%%%%%%%%%%%%%%%%%%%%%
%%%%%%%%%%%%%%%%%%%%%%%%%%%%%%%%%%%%%%%%%%%%%%%%%%%

We have discussed ChPT
with a nonzero $\theta$ term.
As a result of {\it CP} violation, the vacuum of chiral
fields is shifted to a nontrivial element 
on the $SU(N_f)$ group manifold.
We have calculated this vacuum shift at 
${\cal O}(\theta^3)$ level, 
as well as the one-loop corrections,
from which the topological susceptibility and
$c_4$, the coefficient of $\theta^4$ in the QCD
vacuum energy, are extracted.
%In the chiral Lagrangian, this shift can be 
%regarded as a source which 

The {\it CP} violation also causes mixing
among different {\it CP} eigenstates, 
between scalar and pseudoscalar,
or vector and axialvector operators.
We have calculated the mesonic two-point
functions upto ${\cal O}(\theta^2)$ 
to the one-loop order and $\theta$ dependence
of the pion mass and decay constant are obtained.

We also have evaluated the effects of fixing topology, 
by Fourier transform with respect to $\theta$.
We found that the effect of fixing topology is 
considerably suppressed as expected; 
the tree level diagram only affects
on the NLO corrections, one-loop diagram
only contributes to the NNLO corrections, and so on.

As applications  of this study, 
it would be interesting to investigate three or four 
point functions, {\it CP} odd observables as well.
It would be also important to compare our results
with lattice QCD simulations.

\acknowledgments

The authors thank the member of JLQCD and TWQCD Collaborations
for discussions and their encouragement to this study.
In particular SA thanks S. Hashimoto and T. Onogi for discussions.
HF thanks P.~H.~Damgaard for discussions and many useful comments.
The work of SA is supported  in part by Grants-in-Aid of 
the Ministry of Education, Culture, Sports, Science and 
Technology of Japan (Grants No. 20340047, No. 20105001, No. 20105003).
The work of HF is supported by the Grant-in-Aid for Nagoya
University Global COE Program, "Quest for Fundamental Principles in the
Universe: from Particles to the Solar System and the Cosmos," 
from the Ministry of Education, Culture, Sports, Science and Technology of Japan.

%% file: appendix.tex
\appendix
%%%%%%%%%%%%%%%%%%%%%%%%%%%%%%%%%%%%%%%%%%%%%%%%%%%
%%%%%%%%%%%%%%%%%%%%%%%%%%%%%%%%%%%%%%%%%%%%%%%%%%%
%%%%%%%%%%%%%%%%%%%%%%%%%%%%%%%%%%%%%%%%%%%%%%%%%%%
\section{
Explicit expression for $G(x, M_{ii}^2,M_{jj}^2)$
}
\label{app:G-property}
%\setcounter{equation}{0}
%%%%%%%%%%%%%%%%%%%%%%%%%%%%%%%%%%%%%%%%%%%%%%%%%%%
%%%%%%%%%%%%%%%%%%%%%%%%%%%%%%%%%%%%%%%%%%%%%%%%%%%
%%%%%%%%%%%%%%%%%%%%%%%%%%%%%%%%%%%%%%%%%%%%%%%%%%%

The diagonal part of the correlator
%%%% G 
\begin{eqnarray}
%-\langle \xi_{ii}(x)\xi_{jj}(y)\rangle &=& 
G(x-y,M_{ii}^2(\theta),M^2_{jj}(\theta))
%\nonumber\\ &=&
%\frac{1}{V}\sum_{p}\frac{e^{ip(x-y)}}{(p^2+M_{ii}^2)(p^2+M_{jj}^2)
%\left(\sum^{N_f}_f\frac{1}{p^2+M^2_{ff}(\theta)}
%+\frac{N_h}{p^2+M^2_{hh}(\theta)}
%\right)},
\end{eqnarray}
can be, in principle, expressed in terms of $\Delta(x,M^2)$.
In this appendix, we consider the most general case
with arbitrary number of flavors.
The UV divergence of $G$ at $x=0$ is also discussed.
Furthermore, explicit examples for the degenerate theory
and nondegenerate $N_f=N_l+N_s$ flavor case will be given.
The similar discussion can be found in Ref.\cite{Damgaard:2008zs}.
In the following, for simplicity, we omit the argument $\theta$ in $M_{ij}^2(\theta)$.
Therefore, $M_{ij}^2$ means  $M^2_{ij}(\theta)$, unless explicitly stated.

Let us first define a function
%%%%%% Eq f(t) %%%%%%%%%%%%%%%%%%%%%%%%%%%%%%%%%%%%%
\begin{eqnarray}
f(t) \equiv \frac{1}{N_f}\sum^k_i \frac{n_i}{t-M^2_{ii}},
\end{eqnarray}
where $k$ denotes the number of ``different'' quark masses, and
$n_i\geq 1$ is the degeneracy of the $i$th mass, 
which satisfies $\sum^k_in_i=N_f$.
Here, we have ordered the masses $M^2_{ii}< M^2_{i+1\;i+1}$ for all $i$.
Noting that $f(t)$ is a monotonically decreasing function, 
%%%%%% Eq f(t) %%%%%%%%%%%%%%%%%%%%%%%%%%%%%%%%%%%%%
\begin{eqnarray}
\frac{d}{dt}f(t) = -\frac{1}{N_f}
\sum^k_i \frac{n_i}{(t-M^2_{ii})^2}<0,
\end{eqnarray}
and
%%%%%% Eq f(M_ii) %%%%%%%%%%%%%%%%%%%%%%%%%%%%%%%%%%%
\begin{eqnarray}
\lim_{\epsilon \to 0}f(M_{ii}^2+\epsilon)&=&\infty,\;\;\;
\lim_{\epsilon \to 0}f(M_{ii}^2-\epsilon)=-\infty,\\
f(t) & < & 0,\;\;\; \mbox{for}\;\;\; t<M^2_{11} \nonumber \\
f(t) & > & 0, \;\;\; \mbox{for}\;\;\; M^2_{kk}<t,
\end{eqnarray}
one can show that an equation $f(t)=0$ has $k-1$ different
solutions (we denote them by $t=\hat{M}^2_{ii}$), each of them satisfying
%%%%% Eq solution %%%%%%%%%%%%%%%%%%%%%%%%%%%%%%%%%%%
\begin{eqnarray}
M^2_{ii}<\hat{M}^2_{ii}<M^2_{i+1\;i+1},\;\;\;(1\leq i \leq k-1).
\end{eqnarray}
%We illustrate this in the plot Fig.~\ref{fig:f(t)}.

Therefore, $f(-p^2)$ can be written in a rational form: 
%%%%% Eq -f(-p^2) %%%%%%%%%%%%%%%%%%%%%%%%%%%%%%%%%%%
\begin{eqnarray}
-f(-p^2)=\frac{\prod_i^{k-1}(p^2+\hat{M}^2_{ii})}
{\prod_j^{k}(p^2+M^2_{jj})},
\end{eqnarray}
and $G(x,M_{ii}^2,M_{jj}^2)$ can thus be expressed as
%%%%% Eq G by delta %%%%%%%%%%%%%%%%%%%%%%%%%%%%%%%%%%%%%%%%%%%%
\begin{eqnarray}
G(x,M_{ii}^2,M_{jj}^2)&=&\frac{1}{N_fV}\sum_{p}
\frac{e^{ipx}\prod_f^{k}(p^2+M^2_{ff})}
{(p^2+M_{ii}^2)(p^2+M_{jj}^2)\prod_f^{k-1}(p^2+\hat{M}^2_{ff})}
\nonumber\\
&&\hspace{-1in}=\left\{
\begin{array}{cc}
\frac{1}{N_f}\left[ \sum^{k-1}_f A^{(ij)}_f \Delta(x,\hat{M}^2_{ff})
+B^{(ij)}_i \Delta(x,M_{ii}^2) +B^{(ij)}_j \Delta(x,M_{jj}^2)
%C\;\partial_{M^2}\Delta(0,M^2)  
\right] & (M_{ii}^2\neq M_{jj}^2),\\
\frac{1}{N_f}\left[ \sum^{k-1}_f A^{(ii)}_f \Delta(x,\hat{M}^2_{ff})
  +B^{(ii)} \Delta(x,M_{ii}^2) + C^{(ii)} \partial_{M_{ii}^2}\Delta(x,M_{ii}^2)
%C\;\partial_{M^2}\Delta(0,M^2)  
\right] & (M_{ii}^2 =  M_{jj}^2),
\end{array}
\right.\nonumber\\
\end{eqnarray}
where the coefficients $A^{(ij)}_f$'s, {\it etc.} are
given by the residues of 
%%%%% Eq f2 %%%%%%%%%%%%%%%%%%%%%%%%%%%
\begin{eqnarray}
f_2(t)=\frac{\prod_f^{k}(-t+M^2_{ff})}
{(-t+M_{ii}^2)(-t+M_{jj}^2)\prod_f^{k-1}(-t+\hat{M}^2_{ff})},
\end{eqnarray}
[or $-(-t+M_{ii}^2)f_2(t)$ for $C^{(ii)}$], at each pole.
Note that $C^{(ii)}=0$ when $M_{ii}^2$ is equal to
any of the physical masses.

Next, we consider the UV divergence of $G(x=0,M_{ii}^2,M_{jj}^2)$.
%%%%%%% Eq delta divergence %%%%%%%%%%%%%%%%%%%%%%%%
%\begin{eqnarray}
%\Delta(0,M^2) &=& \frac{1}{V}\sum_{p\neq 0}\frac{1}{p^2+M^2}
%\nonumber\\&=&
%\frac{1}{V}\sum_{p\neq 0}\left(\frac{1}{p^2}-
%\frac{M^2}{p^4}\right)+{\cal O}(M^4) .
%\end{eqnarray}
%Here the first term is quadratically divergent, 
%the second term produces a logarithmic divergence, 
%and the remaining ${\cal O}(M^4)$ terms are UV finite.
%As is well-known. the quadratic divergence is absent when we employ 
%dimensional regularization.\\
By expanding the denominator of Eq.(\ref{eq:propG}) in terms of masses, the UV-divergent part of 
$G(0,M_{ii}^2,M_{jj}^2)$ can be written as
%%%%%% Eq G divergence %%%%%%%%%%%%%%%%%%%%%%%%%%%%%
\begin{eqnarray}
\label{eq:Gdiv}
G(0,M_{ii}^2,M_{jj}^2) 
%&=& \frac{1}{V}\sum_{p}
%\frac{1}{(p^2+M_{ii}^2)(p^2+M_{jj}^2)
%\left(\sum_f^{N_f}\frac{1}{p^2+M^2_{ff}}\right)}
%\nonumber\\
&=&
\frac{1}{N_fV}\sum_{p}\left(\frac{1}{p^2}-
\frac{M_{ii}^2+M_{jj}^2}{p^4}+\frac{1}{N_f}\sum^{N_f}_f\frac{M^2_{ff}}{p^4}
%+{\cal O}(M^4)
+\cdots\right)\nonumber\\
%&=&\frac{2}{N_f}\Delta(0,M_{ij}^2)
%-\frac{1}{N_f^2}\sum^{N_f}_f\Delta(0,M^2_{ff})+{\cal O}(M^4)
%\nonumber\\
&=& -\left(\frac{2}{N_f}M_{ij}^2-\frac{1}{N_f^2}\sum^{N_f}_f M^2_{ff}\right)
\left(\frac{1}{\epsilon}+1-\gamma+\ln 4\pi\right)/16\pi^2+\cdots,
\end{eqnarray}
where the logarithmic divergence of the last line is canceled by
a renormalization of $L_i$'s as seen in Sec.~\ref{sec:NLO}.
Note that the quadratic divergence from $1/p^2$ term 
is absent in the dimensional regularization.\\
%As is well-known, the quadratic divergence is absent when we employ 
%dimensional regularization.\\

Here, we give some explicit examples.
For the fully degenerate case, $i.e.$, equal masses 
$M^2_{ff} = M^2_{sea}$ for all sea flavor $f$, 
the above expression for $G$ is greatly simplified,
%%%%% Eq G by delta Nf degenerate %%%%%%%%%%%%%
\begin{eqnarray}
G(x,M_{ii}^2,M_{jj}^2)
&=& 
\left\{
\begin{array}{cc}
\frac{1}{N_f}\left[ 
\frac{M_{ii}^2-M_{sea}^2}{M_{ii}^2-M_{jj}^2}\Delta(x,M_{ii}^2) -
\frac{M_{jj}^2-M_{sea}^2}{M_{ii}^2-M_{jj}^2}\Delta(x,M_{jj}^2) \right]
&  (M_{ii}^2\neq M_{jj}^2),\\
\frac{1}{N_f}\left[ 
\Delta(x,M_{ii}^2) -
(M_{sea}^2-M_{ii}^2)\partial_{M_{ii}^2}\Delta(x,M_{ii}^2) \right]
&  (M_{ii}^2= M_{jj}^2).
\end{array}\right.%\nonumber\\
\end{eqnarray}

For the $N_f=N_l+N_s$ flavor theory,
where we have $N_l$ quarks of mass $m_l$
and $N_s$ quarks of mass $m_s$, 
the equation $f(t)=0$ is easily solved 
and one obtains
%%%%% Eq N_f=2+1 %%%%%%%%%%%%%%%%%%%%%%%%%%%%%%%
\begin{eqnarray}
\label{eq:Gbar2+1}
G(x,M_{ii}^2,M_{jj}^2)
&=& 
\left\{
\begin{array}{cc}
\frac{1}{N_f}\left[
A^{(ij)}  \Delta\left(x,M_\eta^2 \right)
+B^{(ij)}_i \Delta(x,M_{ii}^2) +B^{(ij)}_j \Delta(x,M_{jj}^2) \right] 
&  (M_{ii}^2\neq M_{jj}^2),\\
\frac{1}{N_f}\left[
A^{(ii)}  \Delta\left(x,M_\eta^2\right)
+B^{(ii)} \Delta(x ,M_{ii}^2) +
C^{(ii)}\;\partial_{M^2}\Delta(x,M_{ii}^2)  \right] & (M_{ii}^2= M_{jj}^2),\\
\end{array}\right.\nonumber\\
\end{eqnarray}
where
%%%%% Eq 2+1 alpha beta gamma %%%%%%%%%%%%%%%%%%
\begin{eqnarray}
\label{eq:ABC2+1}
A^{(ij)} &=& \frac{(M^2_{ll}-M^2_\eta)(M^2_{ss}-M^2_\eta)}
{(M_{ii}^2-M_\eta^2)(M_{jj}^2-M_\eta^2)},\;\;\;
B^{(ij)}_i = \frac{(M^2_{ll}-M^2_{ii})(M^2_{ss}-M^2_{ii})}
{(M_{\eta}^2-M_{ii}^2)(M_{jj}^2-M_{ii}^2)},\nonumber\\
B^{(ij)}_j &=& -\frac{(M^2_{ll}-M^2_{jj})(M^2_{ss}-M^2_{jj})}
{(M_{\eta}^2-M_{jj}^2)(M_{jj}^2-M_{ii}^2)},\;\;\;
B^{(ii)} = 1 + \frac{N_l N_s(M_{ll}^2-M_{ss}^2)^2}{N_f^2(M_{ii}^2-M^2_{\eta})^2},
\nonumber\\
C^{(ii)}&=&\frac{(M_{ii}^2-M^2_{ll})(M_{ii}^2-M^2_{ss})}{M_{ii}^2-M^2_{\eta}},
\end{eqnarray}
where $M^2_{ll}=2m_l\Sigma/F^2$, 
$M^2_{ss}=2m_s\Sigma/F^2$ and $M_\eta^2=(N_sM_{ll}^2+N_l M_{ss}^2)/N_f$.

%%%%%%%%%%%%%%%%%%%%%%%%%%%%%%%%%%%%%%%%%%%%%%%%%%%
%%%%%%%%%%%%%%%%%%%%%%%%%%%%%%%%%%%%%%%%%%%%%%%%%%%
%%%%%%%%%%%%%%%%%%%%%%%%%%%%%%%%%%%%%%%%%%%%%%%%%%%
\section{
Correlators of $\xi$'s at finite volume
}
\label{app:Delta}
%\setcounter{equation}{0}
%%%%%%%%%%%%%%%%%%%%%%%%%%%%%%%%%%%%%%%%%%%%%%%%%%%
%%%%%%%%%%%%%%%%%%%%%%%%%%%%%%%%%%%%%%%%%%%%%%%%%%%
%%%%%%%%%%%%%%%%%%%%%%%%%%%%%%%%%%%%%%%%%%%%%%%%%%%

In this appendix, we list several useful formulas for
the correlation functions of $\xi$ fields or 
$\Delta(x,M^2)$ at finite volume $V=L^3 T$.
In particular, we consider the zero-mode projection
or the three-dimensional spatial integrals.

A useful identity is
%%%%%%%%%%%%%%%%%%%%%%%%%%%%%%%%%%%%%%
%%%%%%%%%% Eq summation formula
\begin{eqnarray}
\hspace{-0.25in}
\sum_{n}\frac{g\left(\frac{2\pi n}{L}\right)
e^{i\frac{2\pi n}{L}x}}{\left(\frac{2\pi n}{L}\right)^2+M^2}
=\frac{L}{4M}\frac{1}{\sinh\left(\frac{ML}{2}\right)}
\left[
g(iM)e^{-M(x-L/2)}+g(-iM)e^{M(x-L/2)}
\right],
\end{eqnarray}
which holds for an arbitrary regular function $g(p)$.
%The zero-mode projection of $\Delta(x,M^2)$,
For example, by setting $g=1$, it is easy to obtain
%%%%%%%%%%%%%%%%%%%%%%%%%%%%%%%%%%%%%%
%%%%%%%%%% Eq deltabar x integration
\begin{eqnarray}
\int d^3 x \Delta(x,M^2)
%&=&\frac{L^3}{V}\sum_{n_t\neq 0}\frac{e^{i\frac{2\pi n_t}{T}t}}
%{\left(\frac{2\pi n_t}{T}\right)^2+M^2_{ij}}\nonumber\\
&=&\frac{1}{2M}\frac{\cosh(M(t-T/2))}{\sinh\left(MT/2\right)},\\
\int d^3 x \partial_0\Delta(x,M^2)
%&=&\frac{L^3}{V}\sum_{n_t\neq 0}\frac{e^{i\frac{2\pi n_t}{T}t}}
%{\left(\frac{2\pi n_t}{T}\right)^2+M^2_{ij}}\nonumber\\
&=&\frac{1}{2}\frac{\sinh(M(t-T/2))}{\sinh\left(MT/2\right)},
\end{eqnarray}
where we denote $t=x^0$.

Rather nontrivial ones are
%%%%%%%%% Eq Delta Delta %%%%%%%%%%%%%%
\begin{eqnarray}
C_{\Delta^2}(t, M_1^2,M_2^3) &\equiv&
\int d^3 x \Delta(x,M_1^2)\Delta(x,M_2^2) 
\nonumber\\
%\hspace{-1.8in}
&=&\frac{T}{V}\sum_{\vec{q}=(q_1,q_2,q_3)}
\frac{\cosh(|q^\prime_1|(t-T/2))}{2|q^\prime_1|
\sinh(|q^\prime_1|T/2)}
\frac{\cosh(|q^\prime_2|(t-T/2))}{2|q^\prime_2|
\sinh(|q^\prime_2|T/2)}\\
C_{\partial \Delta\Delta}(t, M_1^2,M_2^3) &\equiv&
\int d^3 x \partial_0\Delta(x,M_1^2)\Delta(x,M_2^2),
\nonumber\\
&=&\frac{T}{V}\sum_{\vec{q}=(q_1,q_2,q_3)}
\frac{\sinh(|q^\prime_1|(t-T/2))}{2
\sinh(|q^\prime_1|T/2)}
\frac{\cosh(|q^\prime_2|(t-T/2))}{2|q^\prime_2|
\sinh(|q^\prime_2|T/2)},\\
C_{\partial \Delta \partial \Delta}(t, M_1^2,M_2^3) &\equiv&
\int d^3 x \partial_0\Delta(x,M_1^2) \partial_0 \Delta(x,M_2^2),
\nonumber\\
&=&\frac{T}{V}\sum_{\vec{q}=(q_1,q_2,q_3)}
\frac{\sinh(|q^\prime_1|(t-T/2))}{2
\sinh(|q^\prime_1|T/2)}
\frac{\sinh(|q^\prime_2|(t-T/2))}{2%|q^\prime_2|
\sinh(|q^\prime_2|T/2)},\\
C_{\Delta\partial^2\Delta}(t, M_1^2,M_2^3) &\equiv&
\int d^3 x \Delta(x,M_1^2)\partial_0^2 \Delta(x,M_2^2),
\nonumber\\
&=&\frac{T}{V}\sum_{\vec{q}=(q_1,q_2,q_3)}
\frac{\cosh(|q^\prime_1|(t-T/2))}{2|q^\prime_1|
\sinh(|q^\prime_1|T/2)}
\frac{|q^\prime_2|\cosh(|q^\prime_2|(t-T/2))}{2%|q^\prime_2|
\sinh(|q^\prime_2|T/2)},
\end{eqnarray}
where $|q^\prime_i|\equiv \sqrt{\vec{q}^2+M_i^2}$.
As one expects, they are ${\cal O}(e^{-(M_1+M_2)t})$ 
for large $T$, which describes 
the two pion state's propagation.

In this paper, we also need
%%%%%%%%% Eq Delta Delta^2 %%%%%%%%%%%%%%%%%%%%
\begin{eqnarray}
%C_{\Delta\Delta^2}(t,M_1^2,M_2,M_3^2)&\equiv&
C_{\Delta\Delta^2}(t,M_1^2; M^2_2,M_3^2)&\equiv&
\int d^3 x \int d^4 y \Delta(x-y,M_1^2)\Delta(y,M_2^2)\Delta(y,M_3^2)
\nonumber\\&=&
%&&\hspace{-2.7in}=
%\frac{1}{T}\sum_{p_0} \frac{e^{ip_0 t}}{p_0^2+M_1^2}\times
%\left[\frac{1}{V}\sum_{p^\prime} \frac{1}{\{(p_0/2+p_0^\prime)^2+\vec{p^\prime}^2+M_2^2\}\{(p_0/2-p_0^\prime)^2+\vec{p^\prime}^2+M_3^2\}}\right]
%\nonumber\\&=&
%&&\hspace{-2.7in}=
%\frac{1}{T}\sum_{p_0} \frac{e^{ip_0 t}}{p_0^2+M_1^2}\times
%\left[\int_0^1 dx \frac{1}{V}\sum_{p^\prime} \frac{1}{\{((x-1/2)p_0+p_0^\prime)^2+\vec{p^\prime}^2+F(p_0,x,M_2^2,M_3^2)\}^2}\right]
%\nonumber\\&=&%\hspace{-1.7in}=
\frac{1}{T}\sum_{p_0} \frac{e^{ip_0 t}}{p_0^2+M_1^2}\times 
\left[\frac{1}{16\pi^2}\left(\frac{1}{\epsilon}+1-\gamma+\ln 4\pi\right)
\right.\nonumber\\&&\left.\hspace{-1in}
+\int_0^1 dx \left(-\frac{\ln F(p^2_0,x,M_2^2,M_3^2)+1}{16\pi^2}+
h^V(x,p_0,F(p^2_0,x,M_2^2,M_3^2))\right)\right],
%\hspace{-2.7in}=
%\frac{1}{2M_{1}}\frac{\cosh(M_{1}(t-T/2))}{\sinh\left(M_{1}T/2\right)}
%\times \left(-\frac{\ln \Lambda^2}{16\pi^2}\right)
\end{eqnarray}
and
\begin{eqnarray}
C_{\partial \Delta\Delta\Delta}(t,M_1^2,M^2_2;M_3^2)&\equiv&
\int d^3 x \int d^4 y \partial_0\Delta(x-y,M_1^2)\Delta(x-y,M_2^2)\Delta(y,M_3^2)
\nonumber\\&=&
\frac{1}{T}\sum_{p_0} \frac{ip_0e^{ip_0 t}}{p_0^2+M_3^2}\times 
\left[\frac{1}{32\pi^2}\left(\frac{1}{\epsilon}+1-\gamma+\ln 4\pi\right)
\right.\nonumber\\&&\left.\hspace{-1.5in}
+\int_0^1 dx (1-x)\left(-\frac{\ln F(p^2_0,x,M_1^2,M_2^2)+1}{16\pi^2}+
h^V(x,p_0,F(p^2_0,x,M_1^2,M_2^2))\right)\right] .
%&&\hspace{-2.7in}=
%\frac{1}{T}\sum_{p_0} \frac{e^{ip_0 t}}{p_0^2+M_3^2}\times
%\left[\frac{1}{V}\sum_{p^\prime} \frac{i(p_0/2+p_0^\prime)}{\{(p_0/2+p_0^\prime)^2+\vec{p^\prime}^2+M_1^2\}\{(p_0/2-p_0^\prime)^2+\vec{p^\prime}^2+M_2^2\}}\right]\nonumber\\
%&&\hspace{-2.7in}=
%\frac{1}{T}\sum_{p_0} \frac{e^{ip_0 t}}{p_0^2+M_3^2}\times
%\left[\int_0^1 dx \frac{1}{V}\sum_{p^\prime} \frac{i((x-1/2)p_0+p_0^\prime)-i(x-1)p_0}{\{((x-1/2)p_0+p_0^\prime)^2+\vec{p^\prime}^2+F(p_0,x,M_1^2,M_2^2)\}^2}\right]\nonumber\\
%&&\hspace{-2.7in}=
%\frac{1}{2}\frac{\partial}{\partial t}\left[
%\frac{1}{2M_{3}}\frac{\cosh(M_{3}(t-T/2))}{\sinh\left(M_{3}T/2\right)}
%\times \left(-\frac{\ln \Lambda^2}{16\pi^2}\right)\right]
%+ C_{\partial \Delta\Delta\Delta}(t,M_1^2,M^2_2;M_3^2),
\end{eqnarray}
Here $F(p^2_0,x,M_2^2,M_3^2)=x(1-x)p_0^2+xM_2^2+(1-x)M_3^2$, and
%%%%%%% C_Delta Delta^2 %%%%%%%%%%%%%%%%%
\begin{eqnarray}
%C_{\Delta\Delta^2}(t,M^2; M^2_2,M_3^2)&\equiv&
%\frac{1}{T}\sum_{p_0} \frac{e^{ip_0 t}}{p_0^2+M^2}\times 
%\left[\int_0^1 dx \left(-\frac{\ln F+1}{16\pi^2}+
%h^V(x,p_0,F)\right)\right],\\
%C_{\partial \Delta\Delta\Delta}(t,M_1^2, M^2_2; M^2)&\equiv&
%\frac{1}{T}\sum_{p_0} \frac{e^{ip_0 t}}{p_0^2+M^2}\times 
%\left[\int_0^1 dx \left(-\frac{\ln F+1}{16\pi^2}+
%h^V(x,p_0,F)\right)\times(-i(x-1)p_0)\right],\nonumber\\\\
h^V(x,p_0,F)&\equiv& \sum_{a\neq 0}\int \frac{d^4p^\prime}{(2\pi)^4}
\frac{e^{-ip^\prime a}}{\{(p_0^\prime+p_0(x-1/2))^2+\vec{p^\prime}^2+F\}^2}
\nonumber\\&=&
 - \sum_{a\neq 0} \cos (p_0(x-1/2) a_0)\frac{\partial}{\partial M^2}
\left.\left(\frac{M}{4\pi^2|a|}K_1(M|a|)\right)\right|_{M=\sqrt{F}}
\end{eqnarray}
where the summation is taken over $a_\mu=n_\mu L_\mu$. 
Note that $h^V(x,p_0,F)$ is exponentially small $\sim {\cal O}(e^{-\sqrt{F}L})$.
Since both of $\ln F(p^2_0,x,M_2^2,M_3^2)$ and $h^V(x,p_0,F)$
are regular with respect to $p_0$ for  nonzero masses
and are symmetric under the flip of $p_0\to -p_0$,
one obtains
%%%%%%%%% Eq Delta Delta^2 %%%%%%%%%%%%%%%%%%%%
\begin{eqnarray}
%C_{\Delta\Delta^2}(t,M_1^2,M_2,M_3^2)&\equiv&
C_{\Delta\Delta^2}(t,M_1^2; M^2_2,M_3^2)&=&
\frac{1}{2M_1}\frac{\cosh(M_1(t-T/2))}{\sinh\left(M_1T/2\right)}
\left[\frac{\left(\frac{1}{\epsilon}+1-\gamma+\ln 4\pi\right)}{16\pi^2}
+h(M_1^2,M_2^2,M_3^2)\right],\nonumber\\
%\right.\nonumber\\&&\left.\hspace{-1in}
%+\int_0^1 dx \left(-\frac{\ln F(-M^2_1,x,M_2^2,M_3^2)+1}{16\pi^2}+
%h^V(x,iM_1,F(-M^2_1,x,M_2^2,M_3^2))\right)\right],
\end{eqnarray}
and
\begin{eqnarray}
C_{\partial \Delta\Delta\Delta}(t,M_1^2,M^2_2;M_3^2)
&=&\frac{\sinh(M_3(t-T/2))}{2\sinh\left(M_3T/2\right)}
\left[\frac{\left(\frac{1}{\epsilon}+1-\gamma+\ln 4\pi\right)}{32\pi^2}
+h^\prime(M_3^2,M_1^2,M_2^2)\right],\nonumber\\
\end{eqnarray}
where
%%%%%%%%%%% Eq h and h^prime %%%%%%%%%%%%%%%%%%%%
\begin{eqnarray}
h(M_1^2,M_2^2,M_3^2) &\equiv& 
\int_0^1 dx \left[-\frac{\ln F(-M_1^2,x,M_2^2,M_3^2)+1}{16\pi^2}
\right.\nonumber\\&&\hspace{-1in}\left.
- \sum_{a\neq 0} \cosh \left(M_1\left(x-\frac{1}{2}\right) a_0\right)
\frac{\partial}{\partial M^2}
\left.\left(\frac{M}{4\pi^2|a|}K_1(M|a|)\right)
\right|_{M=\sqrt{F(-M_1^2,x,M_2^2,M_3^2)}}
\right],\\
h^\prime(M_3^2,M_1^2,M_2^2) &\equiv& 
\int_0^1 dx (1-x)\left[-\frac{\ln F(-M_3^2,x,M_1^2,M_2^2)+1}{16\pi^2}
\right.\nonumber\\&&\hspace{-1in}\left.
- \sum_{a\neq 0} \cosh \left(M_3\left(x-\frac{1}{2}\right) a_0\right)
\frac{\partial}{\partial M^2}
\left.\left(\frac{M}{4\pi^2|a|}K_1(M|a|)\right)
\right|_{M=\sqrt{F(-M_3^2,x,M_1^2,M_2^2)}}
\right].
\end{eqnarray}
Note that the logarithmic divergences are canceled
by the renormalization of $L_i$'s (See Appendix~\ref{app:Corr}).

%%%%%%%%%%%%%%%%%%%%%%%%%%%%%%%%%%%%%%%%%%%%%%%%%%%
%%%%%%%%%%%%%%%%%%%%%%%%%%%%%%%%%%%%%%%%%%%%%%%%%%%
%%%%%%%%%%%%%%%%%%%%%%%%%%%%%%%%%%%%%%%%%%%%%%%%%%%
\section{
Details of two-point correlators 
}
\label{app:Corr}
%\setcounter{equation}{0}
%%%%%%%%%%%%%%%%%%%%%%%%%%%%%%%%%%%%%%%%%%%%%%%%%%%
%%%%%%%%%%%%%%%%%%%%%%%%%%%%%%%%%%%%%%%%%%%%%%%%%%%
%%%%%%%%%%%%%%%%%%%%%%%%%%%%%%%%%%%%%%%%%%%%%%%%%%%

In this appendix, we list several equations
which are needed to calculate
%present the details of
%two-point correlators of axial and 
%pseudoscalar operators given in 
Eqs.(\ref{eq:PP}), (\ref{eq:AP}) and (\ref{eq:AA}).

We first define
%%%%%% Eq X definition %%%%%%%%%%%%%%%%%%
\begin{eqnarray}
X_{ij}(x)&\equiv& 
%\langle \xi^2(x)_{ij}\xi^2(0)_{ji}\rangle
%\nonumber\\\&=&
\sum_f^{N_f}\Delta(x,M_{if}^2)\Delta(x,M_{jf}^2)
%\nonumber\\&&
-\Delta(x,M_{ij}^2)G(x,M_{ii}^2,M_{ii}^2)
\nonumber\\&&
-\Delta(x,M_{ij}^2)G(x,M_{jj}^2,M_{jj}^2)
%\nonumber\\&&
-2\Delta(x,M_{ij}^2)G(x,M_{ii}^2,M_{jj}^2),\\
%%%%%%%%% Y_0 definition 
Y_0^{ij}(x)%\equiv \langle \xi^2(x)_{ij}\xi^2(0)_{ji}\rangle
%\nonumber\\
&\equiv&\sum_f^{N_f}[\partial_0\Delta(x,M_{if}^2)
\Delta(x,M_{jf}^2)
-\partial_0\Delta(x,M_{jf}^2)
\Delta(x,M_{if}^2)]
\nonumber\\&&
-\Delta(x,M_{ij}^2)\partial_0[
G(x,M_{ii}^2,M_{ii}^2)
%\nonumber\\&&
-G(x,M_{jj}^2,M_{jj}^2)]
\nonumber\\&&
+\partial_0\Delta(x,M_{ij}^2)[
G(x,M_{ii}^2,M_{ii}^2)
%\nonumber\\&&
-G(x,M_{jj}^2,M_{jj}^2)],\\
%\nonumber\\\\
%%%%%%%%% W definition 
W_{00}^{ij}(x)%\equiv \langle \xi^2(x)_{ij}\xi^2(0)_{ji}\rangle
%\nonumber\\
&\equiv&\sum_f^{N_f}[\partial^2_0\Delta(x,M_{jf}^2)
\Delta(x,M_{if}^2)
+\partial^2_0\Delta(x,M_{if}^2)
\Delta(x,M_{jf}^2)
%\nonumber\\&&\hspace{1in}
-2\partial_0\Delta(x,M_{if}^2)
\partial_0\Delta(x,M_{jf}^2)]
\nonumber\\&&
+2\partial_0^2\Delta(x,M_{ij}^2)
A(x,M_{ii}^2,M_{jj}^2)
%\nonumber\\&&
%\nonumber\\&&
+2\Delta(x,M_{ij}^2)\partial_0^2
A(x,M_{ii}^2,M_{jj}^2)
\nonumber\\&&
-4\partial_0\Delta(x,M_{ij}^2)\partial_0
A(x,M_{ii}^2,M_{jj}^2).
\end{eqnarray}
In the above expressions, we have omitted
the argument $\theta$ in the mass
unless its difference from $M^2_{ij}(\theta=0)$ is important.
Note that $Y_0^{ij}$ is always finite and
$Y_0^{ij}=0$ when $m_i=m_{j}$.

Using the formulas in Appendix~\ref{app:Delta},
we can perform the zero momentum projection in a finite volume.
It is straightforward to obtain 
%%%%%%%%%% Eq \int d^3 x X %%%%%%%%%%%%%%%%%%%%%%%%%
\begin{eqnarray}
C^{ij}_X(t)  \equiv \int d^3 x X_{ij}(x),\;\;
C^{ij}_{Y_0}(t) \equiv \int d^3 x Y_0^{ij}(x),\;\;\;
C^{ij}_{W_{00}}(t)\equiv \int d^3 x W_{00}^{ij}(x).
\end{eqnarray}
For example, $C^{ij}_X(t)$ is explicitly given by
%(here we again omit the argument for $\theta$ dependence)
%%%%%%%%%% Eq \int d^3 x X %%%%%%%%%%%%%%%%%%%%%%%%%
\begin{eqnarray}
C^{ij}_X(t)  &\equiv& \int d^3 x X_{ij}(x) \nonumber\\
&&\hspace{-0.6in}=
%&=&  
\sum_f^{N_f} C_{\Delta^2}(t, M^2_{if},M^2_{jf})
\nonumber\\&&
\hspace{-0.6in}
- \frac{1}{N_f}\left(\sum^{k-1}_l A^{(ii)}_l C_{\Delta^2}(t, M^2_{ij}, \hat{M}^2_{ll})
+B^{(ii)} C_{\Delta^2}(t, M^2_{ij}, M^2_{ii})
+C^{(ii)} \partial_{M^2}C_{\Delta^2}(t, M^2_{ij}, M^2)|_{M=M_{ii}}
\right)
\nonumber\\&&
\hspace{-0.6in}
- \frac{1}{N_f}\left(\sum^{k-1}_l A^{(jj)}_l C_{\Delta^2}(t, M^2_{ij}, \hat{M}^2_{ll})
+B^{(jj)} C_{\Delta^2}(t, M^2_{ij}, M^2_{jj})
+C^{(jj)} \partial_{M^2}C_{\Delta^2}(t, M^2_{ij}, M^2)|_{M=M_{jj}}
\right)
\nonumber\\&&
\hspace{-0.6in}
- \frac{2}{N_f}\left(\sum^{k-1}_l A^{(ij)}_l C_{\Delta^2}(t, M^2_{ij}, \hat{M}^2_{ll})
+B^{(ij)}_i C_{\Delta^2}(t, M^2_{ij}, M^2_{ii})
+B^{(ij)}_j C_{\Delta^2}(t, M^2_{ij}, M^2_{jj})
\right), 
\end{eqnarray}
\if0
%%%%%%%%%% Eq \int d^3 x Y %%%%%%%%%%%%%%%%%%%%%%%%%
\begin{eqnarray}
C^{ij}_{Y_0}(t)&\equiv& \int d^3 x Y_0^{ij}(x)\nonumber\\
&=&
%\hspace{-0.7in}
  \sum_f^{N_f} 
\left(C_{\partial\Delta\Delta}(t, M^2_{if},M^2_{jf})
-C_{\partial\Delta\Delta}(t, M^2_{jf},M^2_{if})\right)
\nonumber\\&&
%\hspace{-0.8in}
+ \frac{1}{N_f}%\left(
\sum^{k-1}_l (A^{(ii)}_l- A^{(jj)}_l)
\left(C_{\partial\Delta\Delta}(t, M^2_{ij}, \hat{M}^2_{ll})
-C_{\partial\Delta\Delta}(t, \hat{M}^2_{ll},M^2_{ij})\right)
\nonumber\\&&%\left.
+\frac{1}{N_f}B^{(ii)} \left( C_{\partial\Delta\Delta}(t, M^2_{ij}, M^2_{ii})
-C_{\partial\Delta\Delta}(t, M^2_{ii}, M^2_{ij})\right)
\nonumber\\&&
-\frac{1}{N_f}B^{(jj)} \left(
C_{\partial\Delta\Delta}(t, M^2_{ij}, M^2_{jj})
-C_{\partial\Delta\Delta}(t, M^2_{jj}, M^2_{ij})
\right)
\nonumber\\&&%\hspace{-0.8in}
+\frac{1}{N_f}C^{(ii)} 
\partial_{M^2} \left( C_{\partial\Delta\Delta}(t, M^2_{ij}, M^2)
-C_{\partial\Delta\Delta}(t, M^2, M_{ij}^2)
\right)|_{M^2=M^2_{ii}}
\nonumber\\&&
-\frac{1}{N_f}C^{(jj)}\partial_{M^2}\left(
C_{\partial\Delta\Delta}(t, M^2_{ij}, M^2)
-C_{\partial\Delta\Delta}(t, M^2, M^2_{ij})
\right)|_{M^2=M^2_{jj}},
%\nonumber\\
\end{eqnarray}
%%%%%%%%%% Eq W00
\begin{eqnarray}
\label{eq:W00}
C^{ij}_{W_{00}}(t)&\equiv& \int d^3 x W_{00}^{ij}(x) 
\nonumber\\
&=& -\frac{T}{V}
\sum_f^{N_f}k_{00}^s(M_{fv}^2)+{\cal O}(M^2_{vv}-M^2_{v^\prime v^\prime}),
\end{eqnarray}
\fi
where $k$ is the number of different sea quark masses,
and the coefficients $A, B, C$'s and $\hat{M}_{ll}^2$
are given in Appendix~\ref{app:G-property}.

Rather nontrivial integrals are
%%%%%%%%%%% Eq. Delta X %%%%%%%%%%%%%%%%%%%%
\begin{eqnarray}
\label{eq:DX}
\int d^3 x \int d^4 y \Delta(x-y,M_{ij}^2)X_{ij}(y)
%&&
\nonumber\\&&\hspace{-2in}=
%&=& 
\frac{1}{2M_{ij}}\frac{\cosh(M_{ij}(t-T/2))}{\sinh\left(M_{ij}T/2\right)} 
\left[\left(\frac{N_f}{2}-\frac{2}{N_f}\right)
\frac{\frac{1}{\epsilon}+1-\gamma+\ln 4\pi}{8\pi^2}
+H_{ij}(M_{ij}^2)\right],
\end{eqnarray}
%%%%%%%%%%% Eq. Delta X Delta %%%%%%%%%%%%%%%%%%%%
\begin{eqnarray}
\int d^3 x \int d^4 y d^4z \Delta(x-y,M_{ij}^2)X_{ij}(y-z)\Delta(z,M_{ij}^2)
&&\nonumber\\&&\hspace{-3.9in}
 = \left.\left(-\frac{1}{2M}\frac{\partial}{\partial M}\right)
\left\{\frac{1}{2M}\frac{\cosh(M(t-T/2))}{\sinh\left(MT/2\right)} 
\left[\left(\frac{N_f}{2}-\frac{2}{N_f}\right)
\frac{\frac{1}{\epsilon}+1-\gamma+\ln 4\pi}{8\pi^2}
%\left(\frac{1}{\epsilon}+1-\gamma+\ln 4\pi\right)
+H_{ij}(M^2)
\right]\right\}\right|_{M=M_{ij}},
\nonumber\\
\end{eqnarray}
and 
%%%%%%%%%%% Eq. Delta X %%%%%%%%%%%%%%%%%%%%
\begin{eqnarray}
\label{eq:YD}
\int d^3 x \int d^4 y Y_0^{ij}(x-y)\Delta(y,M^2_{ij})
%&&
%\nonumber\\&&\hspace{-2in}=
&=& 
\frac{1}{2}\frac{\sinh(M_{ij}(t-T/2))}{\sinh\left(M_{ij}T/2\right)} 
H^\prime_{ij}(M_{ij}^2),
\end{eqnarray}
where
%%%%%%%%%%%% Eq. H_ij %%%%%%%%%%%%%%%%%%%%%%%
\begin{eqnarray}
H_{ij}(M^2) &\equiv& \sum_f^{N_f}h(M^2,M^2_{if},M^2_{jf})
-\frac{1}{N_f}\left[
\sum_l^{k-1}(A_l^{(ii)}+A_l^{(jj)}+2A_l^{(ij)})h(M^2,M_{ij}^2,\hat{M}^2_{ll})
\right.\nonumber\\&&\left.
+(B^{(ii)}+2B_i^{(ij)})h(M^2,M_{ij}^2,M^2_{ii})
+(B^{(jj)}+2B_j^{(ij)})h(M^2,M_{ij}^2,M^2_{jj})
\right.\nonumber\\&&\left.
+C^{(ii)}\partial_{M_*^2}h(M^2,M_{ij}^2,M_*^2)|_{M_*=M_{ii}}
+C^{(jj)}\partial_{M_*^2}h(M^2,M_{ij}^2,M_*^2)|_{M_*=M_{jj}}
\right],\nonumber\\\\
H^\prime_{ij}(M^2) &\equiv& \sum_f^{N_f}
[h^\prime(M^2,M^2_{if},M^2_{jf})-h^\prime(M^2,M^2_{jf},M^2_{if})]
\nonumber\\&&
+\frac{1}{N_f}\left[
\sum_l^{k-1}(A_l^{(ii)}-A_l^{(jj)})
(h^\prime(M^2,M_{ij}^2,\hat{M}^2_{ll})-h^\prime(M^2,\hat{M}^2_{ll},M^2_{ij}))
\right.\nonumber\\&&\left.
+B^{(ii)}(h^\prime(M^2,M_{ij}^2,M^2_{ii})-h^\prime(M^2,M_{ii}^2,M^2_{ij}))
\right.\nonumber\\&&\left.
-B^{(jj)}(h^\prime(M^2,M_{ij}^2,M^2_{jj})-h^\prime(M^2,M_{jj}^2,M^2_{ij}))
\right.\nonumber\\&&\left.
+C^{(ii)}\partial_{M_*^2}
(h^\prime(M^2,M_{ij}^2,M_*^2)-h^\prime(M^2,M_*^2,M_{ij}^2))|_{M_*=M_{ii}}
\right.\nonumber\\&&\left.
-C^{(jj)}\partial_{M_*^2}
(h^\prime(M^2,M_{ij}^2,M_*^2)-h^\prime(M^2,M_*^2,M_{ij}^2))|_{M_*=M_{jj}}
\right].
\end{eqnarray}
Note that $1/\epsilon$ divergence is canceled by the renormalization
of $L_7$ and $L_8$.
 
Finally we present the overall coefficients
in Eqs.(\ref{eq:PP}), (\ref{eq:AP}) and (\ref{eq:AA}):
%%%%%%%%%% Eq C definitions %%%%%%%%%%%%%%%%%%%%%%
\begin{eqnarray}
[C^{\theta}_{PP}(m_v,m_{v^\prime})]^{\rm 1-loop}&\equiv&
\frac{(\Sigma^{\rm 1-loop}_{vv^\prime}(\theta))^2}{(F Z_F^{vv^\prime}(\theta))^2}
Z_{PP}^{vv^\prime}(\theta),\\{}
[C^{\theta}_{AP}(m_v,m_{v^\prime})]^{\rm 1-loop}&\equiv&
\Sigma^{\rm 1-loop}_{vv^\prime}(\theta)Z_{AP}^{vv^\prime}(\theta),\\{}
[C^{\theta}_{AA}(m_v,m_{v^\prime})]^{\rm 1-loop}&\equiv&
-(F Z_F^{vv^\prime}(\theta))^2Z_{AA}^{vv^\prime}(\theta),
\end{eqnarray}
where we have defined
$\Sigma^{\rm 1-loop}_{vv^\prime}(\theta)\equiv 
\Sigma (Z_M^{vv^\prime}(\theta)Z_F^{vv^\prime}(\theta))^2$
and dimensionless coefficients $Z$'s are given by
%%%%%%%%% Z_PP M^R %%%%%%%%%%%%%%%%%%%
\begin{eqnarray}
%%%%% Z_PP
Z_{PP}^{vv^\prime}(\theta)&\equiv &
%1-\frac{(\theta_v+\theta_{v^\prime})^2}{4}
\frac{1+\cos(\theta_v+\theta_{v^\prime})}{2}
-\frac{\bar{M}^4\theta^2}{2F^2}\left(\left.\frac{\partial}{\partial M^2}
H^r_{vv^\prime}(M^2,\mu_{sub})\right|_{M=M_{vv^\prime}}\right)
\nonumber\\&&
-\frac{\bar{M}^2\theta}{2F^2}
(\theta_v+\theta_{v^\prime})
\left\{32(N_fL^r_7(\mu_{sub})+L^r_8(\mu_{sub}))
+H^r_{vv^\prime}(M^2_{vv^\prime},\mu_{sub}))\right\},\\
%%%%% Z_AP
Z_{AP}^{vv^\prime}(\theta)&\equiv &
%1-\frac{(\theta_v^2+\theta_{v^\prime}^2)}{4}
\frac{\cos \theta_v +\cos \theta_{v^\prime}}{2}
-\frac{\bar{M}^4\theta^2}{2F^2}\left(\left.\frac{\partial}{\partial M^2}
H^r_{vv^\prime}(M^2,\mu_{sub})\right|_{M=M_{vv^\prime}}\right)
\nonumber\\&&
-\frac{\bar{M}^2\theta}{4F^2}
(\theta_v+\theta_{v^\prime})
\left\{32(N_fL^r_7(\mu_{sub})+L^r_8(\mu_{sub}))
+H^r_{vv^\prime}(M^2_{vv^\prime},\mu_{sub}))\right\}
\nonumber\\&&
+\frac{1}{4}(\theta_v-\theta_{v^\prime})
\frac{\bar{M}^2\theta}{F^2}H^\prime_{vv^\prime}(M^2_{vv^\prime}),
\\
Z_{AA}^{vv^\prime}(\theta)&\equiv &
%1-\frac{(\theta_v-\theta_{v^\prime})^2}{4}
\frac{1+\cos(\theta_v-\theta_{v^\prime})}{2}
-\frac{\bar{M}^4\theta^2}{2F^2}\left(\left.\frac{\partial}{\partial M^2}
H^r_{vv^\prime}(M^2,\mu_{sub})\right|_{M=M_{vv^\prime}}\right)\nonumber \\
&&+\frac{1}{2}(\theta_v-\theta_{v^\prime})
\frac{\bar{M}^2\theta}{F^2}H^\prime_{vv^\prime}(M^2_{vv^\prime}),
%Z_X^{vv^\prime}&\equiv & D_X^{vv^\prime}(M_{vv^\prime}(\theta))
%+32(N_fL_7+L_8).
\end{eqnarray}
where $L^r_7(\mu_{sub})$, $L^r_8(\mu_{sub})$ and 
\begin{equation}
H^r_{vv^\prime}(M^2_{vv^\prime},\mu_{sub})\equiv 
H_{vv^\prime}(M^2_{vv^\prime})+\frac{1}{8\pi^2}
\left(\frac{N_f}{2}-\frac{2}{N_f}\right)\ln \mu_{sub}^2,
\end{equation}
represent the renormalized
values at a reference scale $\mu_{sub}$.
In these functions one can ignore the argument $\theta$ 
of the masses  and just put $M_{vv^\prime}^2=M_{vv^\prime}^2(\theta=0)$,
since they only appear in ${\cal O}(\theta^2)\times {\cal O}(p^2)$ terms.

%% file: appendix2.tex
%%%%%%%%%%%%%%%%%%%%%%%%%%%%%%%%%%%%%%%%%%%%%%%%%%%
%%%%%%%%%%%%%%%%%%%%%%%%%%%%%%%%%%%%%%%%%%%%%%%%%%%
\section{$\theta$ derivatives}
\label{app:theta}
%\setcounter{equation}{0}
%%%%%%%%%%%%%%%%%%%%%%%%%%%%%%%%%%%%%%%%%%%%%%%%%%%
%%%%%%%%%%%%%%%%%%%%%%%%%%%%%%%%%%%%%%%%%%%%%%%%%%%
%%%%%%%%%%%%%%%%%%%%%%%%%%%%%%%%%%%%%%%%%%%%%%%%%%%

In this appendix, we list the $\theta$ derivatives
of various quantities, which are needed to
evaluate those in a fixed topological sector
shown in Sec. \ref{sec:fixedQ}.
Here, we evaluate them at NLO for the second derivatives
while at LO for the 4th derivatives.

For the mass matrix, we obtain
%%%%%%%%%% Eq. M^2(2), M^2 (4) %%%%%%%%%%%%%%
\begin{eqnarray}
[M^{(2)}_{ij}]_{LO} &\equiv& 
\left.\frac{\partial^2}{\partial \theta^2} 
M_{ij}(\theta)\right|_{\theta =0}
=-M_{ij}
\left(\frac{\bar{m}^2}{2m_i m_j}\right),\\{}
 [M^{(4)}_{ij}]_{LO} &\equiv&
\left.\frac{\partial^4}{\partial \theta^4} 
M_{ij}(\theta)\right|_{\theta =0}
=\frac{\bar{M}^2}{4M_{ij}}
\left[\frac{\bar{m}^3}{m_i^3}+\frac{\bar{m}^3}{m_j^3}
+24(a_i+a_j)\right]-\frac{3M_{ij}}{4}
\left(\frac{\bar{m}^2}{m_im_J}\right)^2,
\end{eqnarray}
where the argument $\theta=0$ is omitted.

For the one-loop propagator, or the chiral-log terms, we have
%%%%%%%%% Eq Delta(2) %%%%%%%%%%%%%%%%%%%
\begin{eqnarray}
\Delta^{(2)}(0,M_{ij}^2)&\equiv&
-M_{ij}^2\left(\frac{\bar{m}^2}{m_im_j}\right)
\partial_{M^2}\Delta(0,M^2)|_{M=M_{ij}}.
\end{eqnarray}
Using this,
it is straightforward to calculate
%%%%%%%%% Eq G(2) %%%%%%%%%%%%%%%%%%%%%%%
\begin{eqnarray}
G^{(2)}(0,M_{ii}^2,M_{jj}^2) &\equiv& 
\frac{\partial^2}{\partial \theta^2} 
G(0,M_{ii}^2(\theta),M_{jj}^2(\theta)),\\
A^{(2)}(0,M_{ii}^2,M_{jj}^2) &\equiv&
G^{(2)}(0,M_{ii}^2,M_{jj}^2)-\frac{1}{2}
(G^{(2)}(0,M_{ii}^2,M_{ii}^2)+G^{(2)}(0,M_{jj}^2,M_{jj}^2)),
\end{eqnarray}
and
%%%%%%%%% Eq Z_M, Z_ F %%%%%%%%%%%%%%%%%
\begin{eqnarray}
[Z_M^{ij}]^{(2)} &\equiv& \frac{1}{2F^2}
\left[G^{(2)}(0,M_{ii}^2,M_{jj}^2)
+8(L_4-2L_6)\sum_f^{N_f}M_{ff}^2
\left(\frac{\bar{m}^2}{m_f^2}\right)
\right.\nonumber\\&&\left.\hspace{2in}
+8(L_5-2L_8)M^2_{ij}\left(\frac{\bar{m}^2}{m_im_j}\right)
\right],\nonumber\\ \\{}
[Z_F^{ij}]^{(2)} &\equiv& -\frac{1}{2F^2}
\left[\frac{\sum_f^{N_f}(\Delta^{(2)}(0,M_{if}^2)+
\Delta^{(2)}(0,M_{jf}^2))}{2}+A^{(2)}(0,M_{ii}^2,M_{jj}^2)
\right.\nonumber\\&&\left.\hspace{2in}
+8\left(L_4\sum_{f}^{N_f}M_{ff}^2\frac{\bar{m}^2}{m_f^2}
+L_5M_{ij}^2\left(\frac{\bar{m}^2}{m_im_j}\right)\right)\right].
\end{eqnarray}
It is then easy to obtain the second  derivative
of the mass at NLO:
%%%%%%%% Eq M^(2) NLO %%%%%%%%%%%%%%%%%
\begin{eqnarray}
[M^{(2)}_{ij}]_{NLO} &=&
M^{\rm 1-loop}_{ij}(\theta = 0)
\left[-\left(\frac{\bar{m}^2}{2m_i m_j}\right)
+\frac{\bar{M}^2}{2M_{ij}^2}(b_i+b_j)
+\frac{[Z_M^{ij}]^{(2)}}{Z_M^{ij}(\theta=0)}
\right.\nonumber\\&&\left.
-\frac{\bar{M}^4\left\{32(N_fL^r_7(\mu_{sub})+L^r_8(\mu_{sub}))
+H^r_{ij}(M^2_{ij},\mu_{sub})\right\}}{2F^2M^2_{ij}}
\right].
\end{eqnarray}

For the overall coefficients of the correlators, we have
%%%%%%%%% Eq Z_PP etc. %%%%%%%%%%%%%%%%%
\begin{eqnarray}
%%%%%%% Z_PP
[Z^{vv^\prime}_{PP}]_{LO}^{(2)} &\equiv&
-\frac{1}{2}\left(\frac{\bar{m}}{m_v}+\frac{\bar{m}}{m_{v^\prime}}\right)^2,\\{}
[Z^{vv^\prime}_{PP}]_{NLO}^{(2)} &\equiv&
[Z^{vv^\prime}_{PP}]_{LO}^{(2)}+
\left(\frac{\bar{m}}{m_v}+\frac{\bar{m}}{m_{v^\prime}}\right)
(b_v+b_{v^\prime})
%\nonumber\\&&
-\frac{\bar{M}^2}{F^2}
\left[\bar{M}^2\partial_{M^2}H_{vv^\prime}^r(M^2,\mu_{sub})
\right.\nonumber\\&&\left.
+\left(\frac{\bar{m}}{m_v}+\frac{\bar{m}}{m_{v^\prime}}\right)
\left\{32(N_fL_7^r(\mu_{sub})+L_8^r(\mu_{sub}))
+H^r_{vv^\prime}(M_{vv^\prime}^2,\mu_{sub})\right\}\right],\\{}
%%%%%%%% Z_AP
[Z^{vv^\prime}_{AP}]_{LO}^{(2)} &\equiv&
-\frac{1}{2}\left(\frac{\bar{m}^2}{m_v^2}
+\frac{\bar{m}^2}{m_{v^\prime}^2}\right),\\{}
[Z^{vv^\prime}_{AP}]_{NLO}^{(2)} &\equiv&
[Z^{vv^\prime}_{AP}]_{LO}^{(2)}+
\left(\frac{\bar{m}}{m_v}b_v+\frac{\bar{m}}{m_{v^\prime}}b_{v^\prime}\right)
%\nonumber\\&&
-\frac{\bar{M}^2}{F^2}
\left[\bar{M}^2\partial_{M^2}H_{vv^\prime}^r(M^2,\mu_{sub})
\right.\nonumber\\&&\left.
+\frac{1}{2}\left(\frac{\bar{m}}{m_v}+\frac{\bar{m}}{m_{v^\prime}}\right)
\left\{32(N_fL_7^r(\mu_{sub})+L_8^r(\mu_{sub}))
+H^r_{vv^\prime}(M_{vv^\prime}^2,\mu_{sub})\right\}
\right.\nonumber\\&&\left.
-\frac{1}{2}\left(\frac{\bar{m}}{m_v}
-\frac{\bar{m}}{m_{v^\prime}}\right)H_{vv^\prime}^\prime(M_{vv^\prime}^2)\right],\\{}
%%%%%%% Z_AA
[Z^{vv^\prime}_{AA}]_{LO}^{(2)} &\equiv&
-\frac{1}{2}\left(\frac{\bar{m}}{m_v}-\frac{\bar{m}}{m_{v^\prime}}\right)^2,\\{}
[Z^{vv^\prime}_{AA}]_{NLO}^{(2)} &\equiv&
[Z^{vv^\prime}_{AA}]_{LO}^{(2)}+
\left(\frac{\bar{m}}{m_v}-\frac{\bar{m}}{m_{v^\prime}}\right)
(b_v-b_{v^\prime})
\nonumber\\&&
-\frac{\bar{M}^2}{F^2}
\left[\bar{M}^2\partial_{M^2}H_{vv^\prime}^r(M^2,\mu_{sub})
-\left(\frac{\bar{m}}{m_v}
-\frac{\bar{m}}{m_{v^\prime}}\right)H_{vv^\prime}^\prime(M_{vv^\prime}^2)
%\right.\nonumber\\&&\left.
%+\left(\frac{\bar{m}}{m_v}+\frac{\bar{m}}{m_{v^\prime}}\right)
%\left\{32(N_fL_7^r(\mu_{sub})+L_8^r(\mu_{sub}))
%+H^r_{vv^\prime}(M_{vv^\prime}^2,\mu_{sub})\right\}
\right].
\end{eqnarray}

Their 4th derivatives are given by 
\begin{eqnarray}
[Z^{vv^\prime}_{PP}]_{LO}^{(4)} &\equiv&
\frac{1}{2}\left(\frac{\bar{m}}{m_v}+\frac{\bar{m}}{m_{v^\prime}}\right)
\left[\left(\frac{\bar{m}}{m_v}+\frac{\bar{m}}{m_{v^\prime}}\right)^3
+24(a_v+a_{v^\prime})\right],\\{}
[Z^{vv^\prime}_{AP}]_{LO}^{(4)} &\equiv&
\frac{1}{2}\left[\left(\frac{\bar{m}^4}{m_v^4}
+\frac{\bar{m}^4}{m_{v^\prime}^4}\right)
+24\left(\frac{\bar{m}}{m_v}a_v+\frac{\bar{m}}{m_{v^\prime}}a_{v^\prime}\right)
\right],\\{}
[Z^{vv^\prime}_{AA}]_{LO}^{(4)} &\equiv&
\frac{1}{2}\left(\frac{\bar{m}}{m_v}-\frac{\bar{m}}{m_{v^\prime}}\right)
\left[\left(\frac{\bar{m}}{m_v}-\frac{\bar{m}}{m_{v^\prime}}\right)^3
+24(a_v-a_{v^\prime})\right].
\end{eqnarray}

%% file: ChPTtheta.bbl
\begin{thebibliography}{99}


%%%%%% lattice
%\cite{Aubin:2004fs}
\bibitem{Aubin:2004fs}
Latest results from large scale full QCD simulations can be found in\\
%  C.~Aubin {\it et al.}  [MILC Collaboration],
  %``Light pseudoscalar decay constants, quark masses, and low energy  constants
  %from three-flavor lattice QCD,''
%  Phys.\ Rev.\  D {\bf 70}, 114501 (2004)
%  [arXiv:hep-lat/0407028];
  %%CITATION = PHRVA,D70,114501;%%
%\cite{DelDebbio:2006cn}
%\bibitem{DelDebbio:2006cn}
%  L.~Del Debbio, L.~Giusti, M.~Luscher, R.~Petronzio and N.~Tantalo,
  %``QCD with light Wilson quarks on fine lattices (I): first experiences and
  %physics results,''
%  JHEP {\bf 0702}, 056 (2007)
%  [arXiv:hep-lat/0610059];
  %%CITATION = JHEPA,0702,056;%%
%\cite{Boucaud:2007uk}
%\bibitem{Boucaud:2007uk}
  Ph.~Boucaud {\it et al.}  [ETM Collaboration],
  %``Dynamical twisted mass fermions with light quarks,''
  Phys.\ Lett.\  B {\bf 650}, 304 (2007)
  [arXiv:hep-lat/0701012];
  %%CITATION = PHLTA,B650,304;%%
%\cite{Aoki:2008tq}
%\bibitem{Aoki:2008tq}
  S.~Aoki {\it et al.}  [JLQCD Collaboration],
  %``Two-flavor QCD simulation with exact chiral symmetry,''
  Phys.\ Rev.\  D {\bf 78}, 014508 (2008)
  [arXiv:0803.3197 [hep-lat]];
  %%CITATION = PHRVA,D78,014508;%%
%\cite{Allton:2008pn}
%\bibitem{Allton:2008pn}
  C.~Allton {\it et al.}  [RBC-UKQCD Collaboration],
  %``Physical Results from 2+1 Flavor Domain Wall QCD and SU(2) Chiral
  %Perturbation Theory,''
  Phys.\ Rev.\  D {\bf 78}, 114509 (2008)
  [arXiv:0804.0473 [hep-lat]];
  %%CITATION = PHRVA,D78,114509;%%
%\cite{Aoki:2008sm}
%\bibitem{Aoki:2008sm}
  S.~Aoki {\it et al.}  [PACS-CS Collaboration],
  %``2+1 Flavor Lattice QCD toward the Physical Point,''
  Phys.\ Rev.\  D {\bf 79}, 034503 (2009)
  [arXiv:0807.1661 [hep-lat]].
  %%CITATION = PHRVA,D79,034503;%%%\cite{Aoki:2008sm}

%%%%%% ChPT
   
%\cite{Weinberg:1968de}
\bibitem{Weinberg:1968de}
  S.~Weinberg,
  %``Nonlinear realizations of chiral symmetry,''
  Phys.\ Rev.\  {\bf 166}, 1568 (1968).
  %%CITATION = PHRVA,166,1568;%%

%\cite{Gasser:1983yg}
\bibitem{Gasser:1983yg}
  J.~Gasser and H.~Leutwyler,
  %``Chiral Perturbation Theory To One Loop,''
  Annals Phys.\  {\bf 158}, 142 (1984);
  %%CITATION = APNYA,158,142;%%
 %\cite{Gasser:1984gg}
%\bibitem{Gasser:1984gg}
%  J.~Gasser and H.~Leutwyler,
  %``Chiral Perturbation Theory: Expansions In The Mass Of The Strange Quark,''
  Nucl.\ Phys.\  B {\bf 250}, 465 (1985).
  %%CITATION = NUPHA,B250,465;%%

%%%%%%%%%%% theta vacuum physics

\bibitem{Vicari:2009}
For the recent review on the $\theta$ vacuum, see
E. Vicari and H. Panagopoulos, Phys. Rep. 470, 93 (2009).

%\cite{Dashen:1970et}
\bibitem{Dashen:1970et}
  R.~F.~Dashen,
  %``Some features of chiral symmetry breaking,''
  Phys.\ Rev.\  D {\bf 3}, 1879 (1971).
  %%CITATION = PHRVA,D3,1879;%%

%\cite{Witten:1980sp}
\bibitem{Witten:1980sp}
  E.~Witten,
  %``Large N Chiral Dynamics,''
  Annals Phys.\  {\bf 128}, 363 (1980);
  %%CITATION = APNYA,128,363;%%
%\cite{Di Vecchia:1980ve}
%\bibitem{Di Vecchia:1980ve}
  P.~Di Vecchia and G.~Veneziano,
  %``Chiral Dynamics In The Large N Limit,''
  Nucl.\ Phys.\  B {\bf 171}, 253 (1980);
  %%CITATION = NUPHA,B171,253;%%
%\cite{Creutz:1995wf}
%\bibitem{Creutz:1995wf}
  M.~Creutz,
  %``Quark masses and chiral symmetry,''
  Phys.\ Rev.\  D {\bf 52}, 2951 (1995)
  [arXiv:hep-th/9505112];
  %%CITATION = PHRVA,D52,2951;%%
%\cite{Smilga:1998dh}
%\bibitem{Smilga:1998dh}
  A.~V.~Smilga,
  %``{QCD} at Theta approx. pi,''
  Phys.\ Rev.\  D {\bf 59}, 114021 (1999)
  [arXiv:hep-ph/9805214].
  %%CITATION = PHRVA,D59,114021;%%
 
%%%%%%%%%%% NEDM effecitive theories
\bibitem{NEDM}
%\cite{Baluni:1978rf}
%\bibitem{Baluni:1978rf}
  V.~Baluni,
  %``CP Violating Effects In QCD,''
  Phys.\ Rev.\  D {\bf 19}, 2227 (1979);
  %%CITATION = PHRVA,D19,2227;%%
%\cite{Crewther:1979pi}
%\bibitem{Crewther:1979pi}
  R.~J.~Crewther, P.~Di Vecchia, G.~Veneziano and E.~Witten,
  % ``Chiral Estimate Of The Electric Dipole Moment Of The Neutron In Quantum
  %Chromodynamics,''
  Phys.\ Lett.\  B {\bf 88}, 123 (1979)
  [Erratum-ibid.\  B {\bf 91}, 487 (1980)];
  %%CITATION = PHLTA,B88,123;%%
%\cite{Pospelov:1999ha}
%\bibitem{Pospelov:1999ha}
  M.~Pospelov and A.~Ritz,
  %``Theta-Induced Electric Dipole Moment of the Neutron via QCD Sum Rules,''
  Phys.\ Rev.\ Lett.\  {\bf 83}, 2526 (1999)
  [arXiv:hep-ph/9904483];
  %%CITATION = PRLTA,83,2526;%%
%\cite{Borasoy:2000pq}
%\bibitem{Borasoy:2000pq}
  B.~Borasoy,
  %``The electric dipole moment of the neutron in chiral perturbation  theory,''
  Phys.\ Rev.\  D {\bf 61}, 114017 (2000)
  [arXiv:hep-ph/0004011];
  %%CITATION = PHRVA,D61,114017;%%
%\cite{Faccioli:2004jz}
%\bibitem{Faccioli:2004jz}
  P.~Faccioli, D.~Guadagnoli and S.~Simula,
 %  ``The neutron electric dipole moment in the instanton vacuum: Quenched
  %versus unquenched simulations,''
  Phys.\ Rev.\  D {\bf 70}, 074017 (2004)
  [arXiv:hep-ph/0406336].
  %%CITATION = PHRVA,D70,074017;%%


%%%%%%%%%%%% theta vacuum simulations 
\bibitem{Aoki:1990}
S. Aoki, A. Gocksch, Phys. Rev. Lett. {\bf 63}, 1125 (1989); erratum,
{\it ibid.} {\bf 65}, 1172 (1990);
S. Aoki, A. Gocksch, A. V. Manohar, S. R. Sharpe,
Phys. Rev. Lett. {\bf 65}, 1092 (1990).

%\cite{Azcoiti:2002vk}
\bibitem{Azcoiti:2002vk}
%  V.~Azcoiti, G.~Di Carlo, A.~Galante and V.~Laliena,
  %``New proposal for numerical simulations of theta-vacuum like systems,''
%  Phys.\ Rev.\ Lett.\  {\bf 89}, 141601 (2002)
%  [arXiv:hep-lat/0203017];
  %%CITATION = PRLTA,89,141601;%%
%\cite{DelDebbio:2002xa}
%\bibitem{DelDebbio:2002xa}
%  L.~Del Debbio, H.~Panagopoulos and E.~Vicari,
  %``Theta dependence of SU(N) gauge theories,''
%  JHEP {\bf 0208}, 044 (2002)
%  [arXiv:hep-th/0204125];
  %%CITATION = JHEPA,0208,044;%%
%\cite{Fukaya:2004kp}
%\bibitem{Fukaya:2004kp}
%  H.~Fukaya and T.~Onogi,
  %``theta vacuum effects on the chiral condensation and the eta' meson
  %correlators in the two-flavor massive QED(2) on the lattice,''
%  Phys.\ Rev.\  D {\bf 70}, 054508 (2004)
%  [arXiv:hep-lat/0403024];
  %%CITATION = PHRVA,D70,054508;%%
%\cite{Shintani:2005xg}
%\bibitem{Shintani:2005xg}
  E.~Shintani {\it et al.},
  %``Neutron electric dipole moment from lattice QCD,''
  Phys.\ Rev.\  D {\bf 72}, 014504 (2005)
  [arXiv:hep-lat/0505022];
  %%CITATION = PHRVA,D72,014504;%%
F. Berruto, T. Blum, K. Orginos, A. Soni, Phys. Rev. D {\bf 73}, 054509 (2006);
%\cite{Aoki:2008gv}
%\cite{Del Debbio:2006df}
%\bibitem{Del Debbio:2006df}
  L.~Del Debbio, G.~M.~Manca, H.~Panagopoulos, A.~Skouroupathis and E.~Vicari,
  %``theta-dependence of the spectrum of SU(N) gauge theories,''
  JHEP {\bf 0606}, 005 (2006)
  [arXiv:hep-th/0603041];
  %%CITATION = JHEPA,0606,005;%%
%\cite{Shintani:2006xr}
%\bibitem{Shintani:2006xr}
  E.~Shintani {\it et al.},
  %``Neutron electric dipole moment with external electric field method in
  %lattice QCD,''
  Phys.\ Rev.\  D {\bf 75}, 034507 (2007)
  [arXiv:hep-lat/0611032];
  %%CITATION = PHRVA,D75,034507;%%
%\bibitem{Giusti:2007tu}
  L.~Giusti, S.~Petrarca and B.~Taglienti,
  %``Theta dependence of the vacuum energy in the SU(3) gauge theory from the
  %lattice,''
  Phys.\ Rev.\  D {\bf 76}, 094510 (2007)
  [arXiv:0705.2352 [hep-th]];
  %%CITATION = PHRVA,D76,094510;%%
%\bibitem{Aoki:2008gv}
  E.~Shintani, S. ~Aoki , Y. ~Kuramashi,
  Phys.\ Rev.\  D {\bf 78}, 014503 (2008)
  [arXiv:0803.0797[hep-lat]].
  %
%  T.~Izubuchi{\it et al.},
%  PoS(LAT2007), 106(2007) 
%  [arXiv:0802.1470[hep-lat]];
  %
%  R.~Horsley{\it et al.},
  %``The electric dipole moment of the nucleon from simulations at imaginary
  %vacuum angle theta,''
%  arXiv:0808.1428 [hep-lat].
  %%CITATION = ARXIV:0808.1428;%%

%\cite{Fukaya:2004kp}
\bibitem{Fukaya:2004kp}
  H.~Fukaya and T.~Onogi,
  %``theta vacuum effects on the chiral condensation and the eta' meson
  %correlators in the two-flavor massive QED(2) on the lattice,''
  Phys.\ Rev.\  D {\bf 70}, 054508 (2004)
  [arXiv:hep-lat/0403024].
  %%CITATION = PHRVA,D70,054508;%%

%%%%%%%%%%%% fixed Q QCD

%\cite{Brower:2003yx}
\bibitem{Brower:2003yx}
  R.~Brower, S.~Chandrasekharan, J.~W.~Negele and U.~J.~Wiese,
  %``QCD at fixed topology,''
  Phys.\ Lett.\  B {\bf 560}, 64 (2003)
  [arXiv:hep-lat/0302005].
  %%CITATION = PHLTA,B560,64;%%

%\cite{Aoki:2007ka}
\bibitem{Aoki:2007ka}
  S.~Aoki, H.~Fukaya, S.~Hashimoto and T.~Onogi,
  %``Finite volume QCD at fixed topological charge,''
  Phys.\ Rev.\  D {\bf 76}, 054508 (2007)
  [arXiv:0707.0396 [hep-lat]].
  %%CITATION = PHRVA,D76,054508;%%

%%%%%%%%%%%%% overlap operator

%\cite{Neuberger:1997fp}
\bibitem{Neuberger:1997fp}
  H.~Neuberger,
  %``Exactly massless quarks on the lattice,''
  Phys.\ Lett.\ B {\bf 417}, 141 (1998)
  [arXiv:hep-lat/9707022];
  %%CITATION = HEP-LAT 9707022;%%
%\cite{Neuberger:1998wv}
%\bibitem{Neuberger:1998wv}
  H.~Neuberger,
  %``More about exactly massless quarks on the lattice,''
  Phys.\ Lett.\ B {\bf 427}, 353 (1998)
  [arXiv:hep-lat/9801031].
  %%CITATION = HEP-LAT 9801031;%%

%%%%%%%%%%%%%% Q preserving det

%\cite{Izubuchi:2002pq}
\bibitem{Izubuchi:2002pq}
  T.~Izubuchi and C.~Dawson  [RBC Collaboration],
  %``Effects of extra Wilson fermions with large negative mass,''
  Nucl.\ Phys.\ Proc.\ Suppl.\  {\bf 106}, 748 (2002);
  %%CITATION = NUPHZ,106,748;%%
%\cite{Vranas:2006zk}
%\bibitem{Vranas:2006zk}
  P.~M.~Vranas,
  %``Gap domain wall fermions,''
  Phys.\ Rev.\  D {\bf 74}, 034512 (2006)
  [arXiv:hep-lat/0606014];
  %%CITATION = PHRVA,D74,034512;%%
%\cite{Fukaya:2006vs}
%\bibitem{Fukaya:2006vs}
  H.~Fukaya {\it et al.} [JLQCD Collaboration],
  %``Lattice gauge action suppressing near-zero modes of H(W),''
  Phys.\ Rev.\  D {\bf 74}, 094505 (2006)
  [arXiv:hep-lat/0607020];
  %%CITATION = PHRVA,D74,094505;%%
%\cite{Renfrew:2009wu}
%\bibitem{Renfrew:2009wu}
  D.~Renfrew, T.~Blum, N.~Christ, R.~Mawhinney and P.~Vranas,
  %``Controlling Residual Chiral Symmetry Breaking in Domain Wall Fermion
  %Simulations,''
  PoS {\bf LATTICE2008}, 048 (2008)
  [arXiv:0902.2587 [hep-lat]];
  %%CITATION = POSCI,LATTICE2008,048;%%%\cite{Renfrew:2009wu}
%\cite{Bruckmann:2009cv}
%\bibitem{Bruckmann:2009cv}
  F.~Bruckmann, F.~Gruber, K.~Jansen, M.~Marinkovic, C.~Urbach and M.~Wagner,
  %``Comparing topological charge definitions using topology fixing actions,''
  arXiv:0905.2849 [hep-lat].
  %%CITATION = ARXIV:0905.2849;%%

%%%%%%%%%%%%%% JLQCD epsilon-regime
%\cite{Fukaya:2007fb}
\bibitem{Fukaya:2007fb}
  H.~Fukaya {\it et al.}  [JLQCD Collaboration],
  %``Two-flavor lattice QCD simulation in the epsilon-regime with exact chiral
  %symmetry,''
  Phys.\ Rev.\ Lett.\  {\bf 98}, 172001 (2007)
  [arXiv:hep-lat/0702003];
  %%CITATION = PRLTA,98,172001;%%
%\cite{Fukaya:2007yv}
%\bibitem{Fukaya:2007yv}
  H.~Fukaya {\it et al.},
  %``Two-flavor lattice QCD in the epsilon-regime and chiral Random Matrix
  %Theory,''
  Phys.\ Rev.\  D {\bf 76}, 054503 (2007)
  [arXiv:0705.3322 [hep-lat]].
  %%CITATION = PHRVA,D76,054503;%%

%%%%%%%%%%%%%% JLQCD F 
%\cite{Fukaya:2007pn}
\bibitem{Fukaya:2007pn}
  H.~Fukaya {\it et al.}  [JLQCD collaboration],
  %``Lattice study of meson correlators in the epsilon-regime of two-flavor
  %QCD,''
  Phys.\ Rev.\  D {\bf 77}, 074503 (2008)
  [arXiv:0711.4965 [hep-lat]];
  %%CITATION = PHRVA,D77,074503;%%%\cite{Noaki:2008iy}
%\bibitem{Noaki:2008iy}
  J.~Noaki {\it et al.}  [JLQCD and TWQCD Collaborations],
  %``Convergence of the chiral expansion in two-flavor lattice QCD,''
  Phys.\ Rev.\ Lett.\  {\bf 101}, 202004 (2008)
  [arXiv:0806.0894 [hep-lat]].
  %%CITATION = PRLTA,101,202004;%%

%%%%%%%%%%%%%%% chi_t in JLQCD

%\cite{Aoki:2007pw}
\bibitem{Aoki:2007pw}
  S.~Aoki {\it et al.}  [JLQCD and TWQCD Collaborations],
  %``Topological susceptibility in two-flavor lattice QCD with exact chiral
  %symmetry,''
  Phys.\ Lett.\  B {\bf 665}, 294 (2008)
  [arXiv:0710.1130 [hep-lat]].
  %%CITATION = PHLTA,B665,294;%%

%%%%%%%%% e-regime %%%%%%%%%%%%%%%%%%%%
\bibitem{e-regime}
%\cite{Hansen:1990un}
%\bibitem{Hansen:1990un}
  F.~C.~Hansen,
  %``FINITE SIZE EFFECTS IN SPONTANEOUSLY BROKEN SU(N) x SU(N) THEORIES,''
  Nucl.\ Phys.\  B {\bf 345}, 685 (1990);
  %%CITATION = NUPHA,B345,685;%%
%\cite{Hansen:1990yg}
%\bibitem{Hansen:1990yg}
  F.~C.~Hansen and H.~Leutwyler,
  %``Charge correlations and topological susceptibility in QCD,''
  Nucl.\ Phys.\  B {\bf 350}, 201 (1991);
  %%CITATION = NUPHA,B350,201;%%
%\cite{Leutwyler:1992yt}
%\bibitem{Leutwyler:1992yt}
  H.~Leutwyler and A.~V.~Smilga,
  %``Spectrum of Dirac operator and role of winding number in QCD,''
  Phys.\ Rev.\  D {\bf 46}, 5607 (1992);
  %%CITATION = PHRVA,D46,5607;%%
%\cite{Splittorff:2002eb}
%\bibitem{Splittorff:2002eb}
  K.~Splittorff and J.~J.~M.~Verbaarschot,
  %``Replica limit of the Toda lattice equation,''
  Phys.\ Rev.\ Lett.\  {\bf 90}, 041601 (2003)
  [arXiv:cond-mat/0209594];
  %%CITATION = PRLTA,90,041601;%%
%\cite{Fyodorov:2002wq}
%\bibitem{Fyodorov:2002wq}
  Y.~V.~Fyodorov and G.~Akemann,
  % ``On the supersymmetric partition function in QCD inspired random matrix
  %models,''
  JETP Lett.\  {\bf 77}, 438 (2003)
%  [Pisma Zh.\ Eksp.\ Teor.\ Fiz.\  {\bf 77}, 513 (2003)]
  [arXiv:cond-mat/0210647];
  %%CITATION = ZFPRA,77,513;%%
%\cite{Splittorff:2003cu}
%\bibitem{Splittorff:2003cu}
  K.~Splittorff and J.~J.~M.~Verbaarschot,
  % ``Factorization of correlation functions and the replica limit of the  Toda
  %lattice equation,''
  Nucl.\ Phys.\  B {\bf 683}, 467 (2004)
  [arXiv:hep-th/0310271].
  %%CITATION = NUPHA,B683,467;%%


%%%%%%%%%%%%%%% PQChPT


%\cite{Sharpe:1997by}
\bibitem{Sharpe:1997by}
  S.~R.~Sharpe,
  %``Enhanced chiral logarithms in partially quenched QCD,''
  Phys.\ Rev.\  D {\bf 56}, 7052 (1997)
  [Erratum-ibid.\  D {\bf 62}, 099901 (2000)]
  [arXiv:hep-lat/9707018];
  %%CITATION = PHRVA,D56,7052;%%
%\cite{Golterman:1997st}
%\bibitem{Golterman:1997st}
  M.~F.~L.~Golterman and K.~C.~L.~Leung,
  %``Applications of Partially Quenched Chiral Perturbation Theory,''
  Phys.\ Rev.\  D {\bf 57}, 5703 (1998)
  [arXiv:hep-lat/9711033];
  %%CITATION = PHRVA,D57,5703;%%
%\cite{Damgaard:2000gh}
%\bibitem{Damgaard:2000gh}
  P.~H.~Damgaard and K.~Splittorff,
  %``Partially quenched chiral perturbation theory and the replica method,''
  Phys.\ Rev.\  D {\bf 62}, 054509 (2000)
  [arXiv:hep-lat/0003017];
  %%CITATION = PHRVA,D62,054509;%%
%\cite{Sharpe:2001fh}
%\bibitem{Sharpe:2001fh}
  S.~R.~Sharpe and N.~Shoresh,
  %``Partially quenched chiral perturbation theory without $\Phi_0$,''
  Phys.\ Rev.\  D {\bf 64}, 114510 (2001)
  [arXiv:hep-lat/0108003];
  %%CITATION = PHRVA,D64,114510;%%
%\cite{Bijnens:2007si}
%\cite{Bijnens:2005ae}
%\bibitem{Bijnens:2005ae}
  J.~Bijnens and T.~A.~Lahde,
  %``Decay constants of pseudoscalar mesons to two loops in three-flavor
  %partially quenched chiPT,''
  Phys.\ Rev.\  D {\bf 71}, 094502 (2005)
  [arXiv:hep-lat/0501014];
  %%CITATION = PHRVA,D71,094502;%%
%\cite{Bijnens:2006jv}
%\bibitem{Bijnens:2006jv}
  J.~Bijnens, N.~Danielsson and T.~A.~Lahde,
  %``Three-flavor partially quenched chiral perturbation theory at NNLO for
  %meson masses and decay constants,''
  Phys.\ Rev.\  D {\bf 73}, 074509 (2006)
  [arXiv:hep-lat/0602003].
  %%CITATION = PHRVA,D73,074509;%%

%%%%%%% two flavor vacuum energy %%%%%%%%%%%%%%%%%%%

%\cite{Lenaghan:2001ur}
\bibitem{Lenaghan:2001ur}
  J.~Lenaghan and T.~Wilke,
  %``Mesoscopic QCD and the Theta vacua,''
  Nucl.\ Phys.\  B {\bf 624}, 253 (2002)
  [arXiv:hep-th/0108166].
  %%CITATION = NUPHA,B624,253;%%

%\cite{Mao:2009sy}
\bibitem{Mao:2009sy}
  Y.~Y.~Mao and T.~W.~Chiu  [TWQCD Collaboration],
  %``Topological Susceptibility to the One-Loop Order in Chiral Perturbation
  %Theory,''
  Phys.\ Rev.\  D {\bf 80}, 034502 (2009)
  [arXiv:0903.2146 [hep-lat]].
  %%CITATION = PHRVA,D80,034502;%%

%%%%%%%% WZW term %%%%%%%%%%%%%%%%%%%%

%\cite{Wess:1971yu}
\bibitem{Wess:1971yu}
  J.~Wess and B.~Zumino,
  %``Consequences of anomalous Ward identities,''
  Phys.\ Lett.\  B {\bf 37}, 95 (1971);
  %%CITATION = PHLTA,B37,95;%%
%\cite{Witten:1983tw}
%\bibitem{Witten:1983tw}
  E.~Witten,
  %``Global Aspects Of Current Algebra,''
  Nucl.\ Phys.\  B {\bf 223}, 422 (1983).
  %%CITATION = NUPHA,B223,422;%%

%%%%%%%% g_1 %%%%%%%%%%%%%%%%%%%%%%%%%%

%\cite{Bernard:2001yj}
\bibitem{Bernard:2001yj}
  C.~Bernard  [MILC Collaboration],
  %``Chiral Logs in the Presence of Staggered Flavor Symmetry Breaking,''
  Phys.\ Rev.\  D {\bf 65}, 054031 (2002)
  [arXiv:hep-lat/0111051].
  %%CITATION = PHRVA,D65,054031;%%



%%%%%%% 2-loop ChPT

%\cite{Bijnens:2004hk}
\bibitem{Bijnens:2004hk}
  J.~Bijnens, N.~Danielsson and T.~A.~Lahde,
  % ``The pseudoscalar meson mass to two loops in three-flavor partially
  %quenched chiPT,''
  Phys.\ Rev.\  D {\bf 70}, 111503 (2004)
  [arXiv:hep-lat/0406017].
  %%CITATION = PHRVA,D70,111503;%%
%\cite{Bijnens:2006jv}
\bibitem{Bijnens:2006jv}
  J.~Bijnens, N.~Danielsson and T.~A.~Lahde,
  % ``Three-flavor partially quenched chiral perturbation theory at NNLO for
  %meson masses and decay constants,''
  Phys.\ Rev.\  D {\bf 73}, 074509 (2006)
  [arXiv:hep-lat/0602003].
  %%CITATION = PHRVA,D73,074509;%%
%\cite{Bijnens:1999hw}
\bibitem{Bijnens:1999hw}
  J.~Bijnens, G.~Colangelo and G.~Ecker,
  %``Renormalization of chiral perturbation theory to order p**6,''
  Annals Phys.\  {\bf 280}, 100 (2000)
  [arXiv:hep-ph/9907333].
  %%C%\cite{Colangelo:2006mp}
\bibitem{Colangelo:2006mp}
  G.~Colangelo and C.~Haefeli,
  %``Finite volume effects for the pion mass at two loops,''
  Nucl.\ Phys.\  B {\bf 744}, 14 (2006)
  [arXiv:hep-lat/0602017].
  %%CITATION = NUPHA,B744,14;%%ITATION = APNYA,280,100;%%
%\cite{Amoros:1999dp}
\bibitem{Amoros:1999dp}
  G.~Amoros, J.~Bijnens and P.~Talavera,
  % ``Two-point functions at two loops in three flavour chiral perturbation
  %theory,''
  Nucl.\ Phys.\  B {\bf 568}, 319 (2000)
  [arXiv:hep-ph/9907264].
  %%CITATION = NUPHA,B568,319;%%
%\cite{Golowich:1997zs}
\bibitem{Golowich:1997zs}
  E.~Golowich and J.~Kambor,
  % ``Two-loop analysis of axialvector current propagators in chiral
  %perturbation theory,''
  Phys.\ Rev.\  D {\bf 58}, 036004 (1998)
  [arXiv:hep-ph/9710214].
  %%CITATION = PHRVA,D58,036004;%%
%\cite{Amoros:2001cp}
\bibitem{Amoros:2001cp}
  G.~Amoros, J.~Bijnens and P.~Talavera,
  % ``QCD isospin breaking in meson masses, decay constants and quark mass
  %ratios,''
  Nucl.\ Phys.\  B {\bf 602}, 87 (2001)
  [arXiv:hep-ph/0101127].
  %%CITATION = NUPHA,B602,87;%%
%\cite{Kaiser:2007kf}
\bibitem{Kaiser:2007kf}
  R.~Kaiser,
  %``On the two-loop contributions to the pion mass,''
  JHEP {\bf 0709}, 065 (2007)
  [arXiv:0707.2277 [hep-ph]].
  %%CITATION = JHEPA,0709,065;%%

%%%%%%%% G %%%%%%%%%%%%%%%%%%%%%%%%%%%%%%

%\cite{Damgaard:2008zs}
\bibitem{Damgaard:2008zs}
  P.~H.~Damgaard and H.~Fukaya,
  %``The Chiral Condensate in a Finite Volume,''
  JHEP {\bf 0901}, 052 (2009)
  [arXiv:0812.2797 [hep-lat]].
  %%CITATION = JHEPA,0901,052;%%

\end{thebibliography}
